\documentclass{astlb}

\usepackage{graphicx}
\usepackage{natbib}
\usepackage{color}

\usepackage[hyperfootnotes=false]{hyperref}

\usepackage{apjfonts}

\usepackage{multirow}

\definecolor{darkblue}{rgb}{0,0,0.9}

\def\smfigure#1#2#3{
  \begin{minipage}{1.0\columnwidth}
    \begin{minipage}{0.049\columnwidth}
      \rotatebox{90}{\small\phantom{0000}#3}
    \end{minipage}
    \begin{minipage}{0.9\columnwidth}
      \includegraphics[bb=40 188 556 678,width=0.97\columnwidth]{#1}
      \centerline{\small #2}
    \end{minipage}

    \vskip 3pt
    ~
  \end{minipage}
}

\def\neff{\ensuremath{N_{\mathit{eff}}}}

\begin{document}

\journalinfo{2012}{38}{0}{1}[0]

\title{Cosmological parameters constraints from galaxy cluster mass
  function measurements in combination with other cosmological data}

\author{R.~A.~Burenin\email{rodion@hea.iki.rssi.ru}\address{1},
  A.~A.~Vikhlinin\address{1,2}
  \addresstext{1}{Space Research Institute (IKI), Moscow}
  \addresstext{2}{Harvard-Smithsonian Center for Astrophysics (CfA),
    Boston, USA} 
}

\shortauthor{Burenin, Vikhlinin}

\shorttitle{Cosmological parameters constraints from galaxy cluster
  mass function}

%\submitted{\today}
\submitted{Dec. 1, 2011}

\begin{abstract}
  We present the cosmological parameters constraints obtained from the
  combination of galaxy cluster mass function measurements (Vikhlinin
  et al., 2009a,b) with new cosmological data obtained during last
  three years: updated measurements of cosmic microwave background
  anisotropy with Wilkinson Microwave Anisotropy Probe (WMAP)
  observatory, and at smaller angular scales with South Pole Telescope
  (SPT), new Hubble constant measurements, baryon acoustic
  oscillations and supernovae Type Ia observations.
  
  New constraints on total neutrino mass and effective number of
  neutrino species are obtained. In models with free number of massive
  neutrinos the constraints on these parameters are notably less
  strong, and all considered cosmological data are consistent with
  non-zero total neutrino mass $\Sigma m_\nu\approx 0.4$\,eV and larger
  than standard effective number of neutrino species, $\neff\approx
  4$. These constraints are compared to the results of neutrino
  oscillations searches at short baselines.

  The updated dark energy equation of state parameters constraints are
  presented. We show that taking in account systematic uncertainties,
  current cluster mass function data provide similarly powerful
  constraints on dark energy equation of state, as compared to the
  constraints from supernovae Type Ia observations.


\end{abstract}

\section{Introduction}
\label{sec:intro} 

The measurements of galaxy cluster mass function give one of the most
sensitive method to measure the cosmological parameters, in particular
the parameters of dark energy equation of state
\citep[e.g.,][]{starobinsky98,haiman01,veller02,wang04,majmohr04}.
Current observational data provide measurements of cluster mass
function accurate enough to obtain powerful constraints on the
parameters of cosmological model
\citep[e.g.,][]{borgani01,henry00,henry04,reipbohr02,
  av03,schuecker03,avav04,mantz08,mantz10a,av09b,vaderlinde10,av10}.

The strongest to date constraints on cosmological parameters using the
data on measurements of galaxy cluster mass function were obtained in
\cite{av09b}. These constraints appears to be similarly powerful as
compared to the constraints from most recent cosmic microwave
background (CMB) measurements made by WMAP, observations of baryon
acoustic oscillations and supernovae Type Ia. These new constraints
are independent and have different cosmological parameter
degeneracies. In this work the confirmation of the existence of the
dark energy was, for the first time, obtained using new independent
method, based on the measurements of the large scale structure growth
rate, not only on the measurements of the geometry of Universe. The
joint analysis with other cosmological data allowed to significantly
improve the measurements of the dark energy equation of state
parameters and also to improve constraints on other cosmological
parameters, e.g. total mass of light neutrinos.

In the last years a significant amount of new cosmological data were
published. These include seven-year data of CMB observations with WMAP
\citep{larson11}, and CMB observations at smaller angular scales
\citep{reichardt09,brown09,dunkley10,keisler11}. Using the
calibration of supernovae Ia absolute luminosities the measurement of
Hubble constant was significantly improved
\citep{riess09,riess11}. Also the new data on observations of
supernovae Ia \citep{hicken09,amanullah10}, and baryon acoustic
oscillations \citep{percival10} were obtained. Cosmological
constraints from the joint analysis of these and others cosmological
data were presented in \cite{komatsu11}, \cite{keisler11} and other
works.

In this paper we present the cosmological parameters constraints,
obtained in result of joint analysis of galaxy cluster mass function
measurements \citep{av09a,av09b} and recent cosmological data
discussed above. As compared to \cite{av09b}, wider set of
cosmological parameters is considered. To calculate joint likelihood
functions in cosmological parameter space we used Markov Chain Monte
Carlo technique \citep[see, e.g.,][]{lewis02}. In order to include
galaxy cluster mass function cosmological data in these calculations,
correspondent software was developed, which we provide for public use.

In our work the new significant constraints on total neutrino mass and
effective number of neutrino species are presented. In order to test
the possibility of the existence of light sterile neutrinos with
masses near $1$\,eV, which were suggested to explain the results of
\emph{LSDN} \citep{aguilar01} and \emph{MiniBooNE}
\citep{aguilararevalo10} experiments, and also recently discovered
so-called reactor neutrino anomaly \citep{mueller11,mention11}, we
consider the models with free number of massive neutrinos. Also in
this paper we give the updated constraints on the parameters of the
dark energy equation of state. We show, that taking in account
systematic uncertainties, dark energy equation of state constraints
based on the existing cluster mass function data and supernovae Type
Ia observations are comparably powerful.

% existing cluster mass funstion data provide
%comparably powerful constraints on dark energy equation of state, as
%compared to the constraints from supernovae Type Ia observations.

\section{Cosmological data}
\label{sec:data}

In our work the data on cluster mass function measurements are taken
with no changes from \cite{av09b}. In this work cluster mass function
was measured using a sample of 86 massive galaxy clusters with masses
measured using \emph{Chandra} X-ray observations with about $10\%$
accuracy \citep{av09a}. Distant ($z\approx0.4$--$0.9$) clusters in
this sample were selected in 400 square degree X-ray galaxy cluster
survey, based on \emph{ROSAT} PSPC pointed data \citep{400d}. Clusters
in local Universe ($z<0.2$) were selected from \emph{ROSAT} All Sky
Survey \citep[see details in][]{av09a}. Confidence regions for
different cosmological parameters were obtained from the analysis of
likelihood function, which was calculated on a grid of cosmological
parameters. The results of these calculations are available on
WWW\footnote{http://hea.iki.rssi.ru/400d/cosm/}. This cosmological
data set is designated below as \emph{CL}.

In our work we used also new cosmological data, which were published
after the issue of paper by \cite{av09b}. The most significant
improvement was achieved in the measurement of Hubble constant, due to
the calibration of supernovae Ia absolute magnitudes, which was made
using the Cepheid observations in SNe Ia host galaxies. After that
from the observations of nearby SNe Ia the measurement $H_0=73.8\pm
2.4$~km\,s$^{-1}$Mpc$^{-1}$ was obtained \citep{riess11}. These
cosmological data are designated below as \emph{$H_0$}. This is much
better measurement as compared to $H_0=72\pm 8$~km\,s$^{-1}$Mpc$^{-1}$
\citep{freedman01} used by \cite{av09b}, and also as compared to the
measurement $H_0=74.2\pm 3.6$~km\,s$^{-1}$Mpc$^{-1}$ \citep{riess09}
used by \cite{komatsu11}.

In our work we used seven-year data of CMB observations from WMAP
\citep{larson11}. For these data likelihood were calculated using the
software from \emph{LAMBDA} archive, version
\emph{4.1}\footnote{http://lambda.gsfc.nasa.gov/product/map/dr4/likelihood\_get.cfm}. Theoretical
CMB anisotropy spectra were calculated using CAMB software, version
January 2011, where recombination was calculated using RECFAST
software \citep{seager99}, version \emph{1.5}, which incorporate a
number of improvements in hydrogen and helium recombination theory,
obtained during the last years (e.g.,
\citealt{dubrovich05,chlubaRA06,chlubaRA07,chlubaRA09,chlubaRA10,RAchluba09,kholupenko07,kholupenko08}). This
cosmological data set is designated as \emph{WMAP7}.

Also we used recently published data on CMB anisotropy at smaller
angular scales obtained with South Pole Telescope \citep[\emph{SPT},
][]{keisler11}. The data were taken from the web page of the
project\footnote{http://pole.uchicago.edu/public/data/keisler11/}.
The contributions of ``poisson'' and ``clustered'' sources were added
to the theoretical CMB anisotropy power spectra according to the
directions in \S 4.1 of paper by \cite{keisler11}. The templates for
these components were also taken from the project web page.  This data
set is designated below as \emph{SPT}.

In addition we used the data on baryon acoustic oscillation
measurements from the data Sloan Digital Sky Survey, Data Release 7,
and the data of \emph{2dF GRS} survey. These data were taken as
Gaussian priors for the ratio of sound horizon at the baryon-drag
epoch to effective distance measure for two redshifts $z=0.2$ and
$z=0.35$ \citep{percival10,komatsu11}. This data set is designated as
\emph{BAO}. In order to study the dark energy equation of state we
also used the data on SNe Ia observations, the
\emph{UNION2}\footnote{http://supernova.lbl.gov/Union/} compilation
\citep{amanullah10}. This data set is designated below as \emph{SN}.

\section{Likelihood Calculation}
\label{sec:mcmc}

In \cite{av09b} likelihood functions for galaxy cluster mass function
data were calculated at cosmological parameters grids, since in that
case these calculations could be performed very effectively. In order
to use these likelihood calculations in Markov Chains Monte Carlo
simulations in multidimensional cosmological parameters space
\citep{lewis02}, we wrote a module for
\texttt{CosmoMC}\footnote{http://cosmologist.info/cosmomc/} software,
where likelihood in arbitrary point of parameters space is calculated
using simple interpolation on parameters grids. In spite of
ideological simplicity, the realization turn out be somewhat
cumbersome due to many dimension grids. Therefore, we think it may be
useful to provide this module for public use. This software can be
found at web page of \emph{CCCP}
project\footnote{http://hea.iki.rssi.ru/400d/cosm/}.

Using this module one can easily include likelihood for galaxy cluster
mass function from \cite{av09b} in calculation of Markov Chains with
\texttt{CosmoMC} software. Also one can use this module to include
this likelihood into existing parameters chains using priors
adjustment procedure. In our work below we mainly used our own
calculations of Markov Chains with \texttt{CosmoMC} software. In some
cases we also used chains, version \emph{WMAP7.2}, taken from
\emph{LAMBDA}
archive\footnote{http://lambda.gsfc.nasa.gov/product/map/dr4/parameters.cfm}.

The constraints for some cosmological parameters in various
cosmological models are presented in Table~\ref{tab:pars}. More
extended information on cosmological constraints obtained in frames of
our work can be found at correspondent web
page\footnote{http://hea.iki.rssi.ru/400d/cosm/combined/}.

In all Figures below we give contours for 68\% and 95\% confidence
regions. In most Figures the pair of contours of larger size
corresponds to the set of cosmological data without the data on
cluster mass function, the pair of contours of smaller size --- with
cluster mass function data taken in account. All confidence intervals
are given at 68\% confidence level, all upper limits --- at 95\%
confidence level.

\section{Systematic uncertainties}
\label{sec:syserr}

Systematic uncertainties of galaxy cluster mass function measurements
were studied in detail by \cite{av09b}. They were not included in
likelihood functions, calculated in this work. Thus, systematical
errors for cosmological parameters constraints should be estimated
separately. These uncertainties can be estimated by measuring the
shifts of likelihood maximum due to variation of observables by their
systematical errors.

For galaxy cluster mass function data the main source of systematical
errors is the uncertainty in cluster mass measurements. The mass
measurements used in our work are based on hot intracluster gas
temperatures and masses and were calibrated using hydrostatic
measurements of their total gravitational masses in nearby clusters
\citep{av09a}. Systematical error of hydrostatic mass measurements was
estimated as $\delta M/M \approx 0.09$ from the comparison with the
cluster mass measurements using weak lensing data taken from
\cite{hoekstra07} and \cite{zhang08}.

\begin{figure}
  \centering
  \begin{minipage}{0.48\linewidth}
    \smfigure{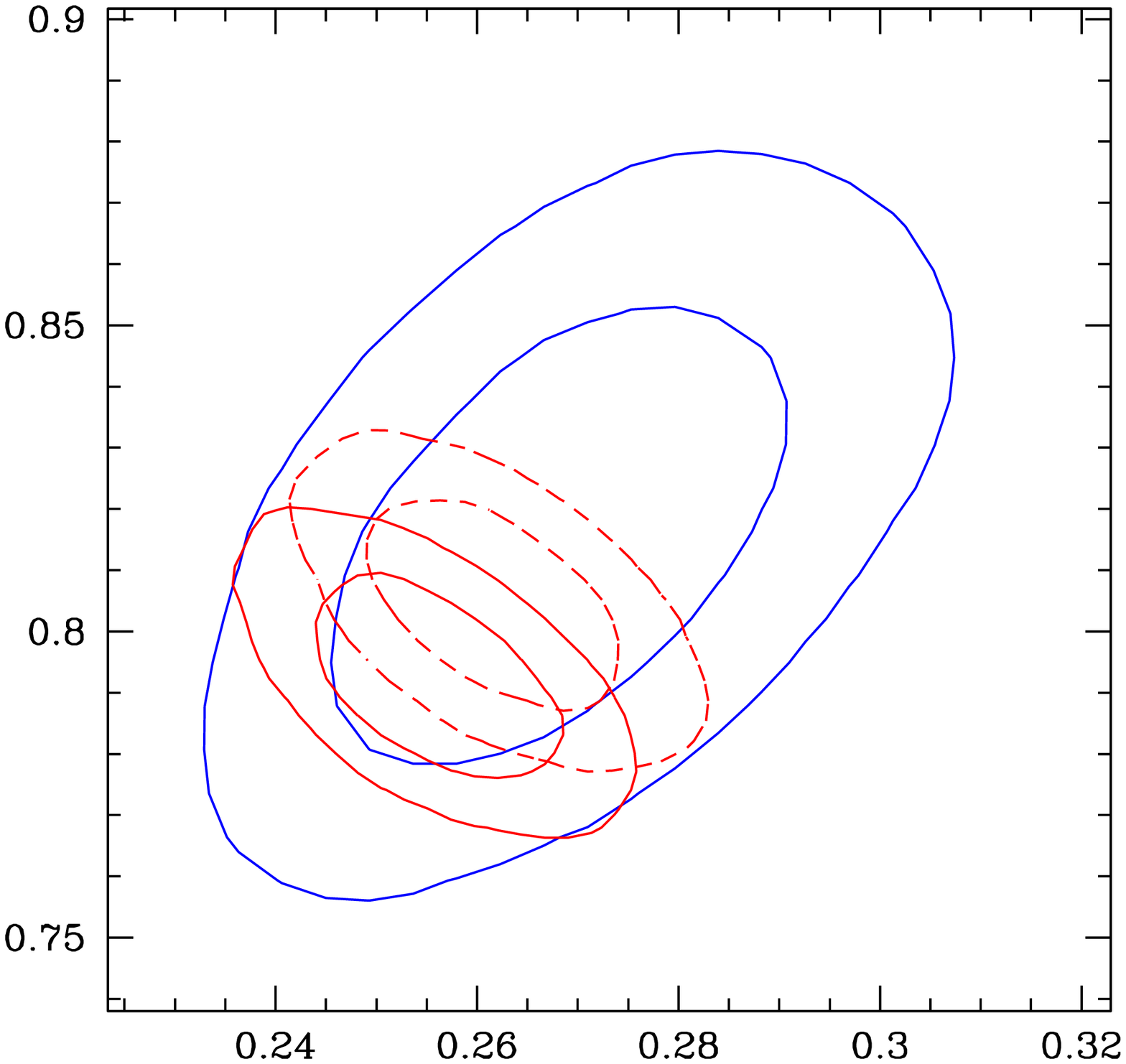}{$\Omega_m$}{$\sigma_8$}
  \end{minipage}
  \begin{minipage}{0.48\linewidth}
    \smfigure{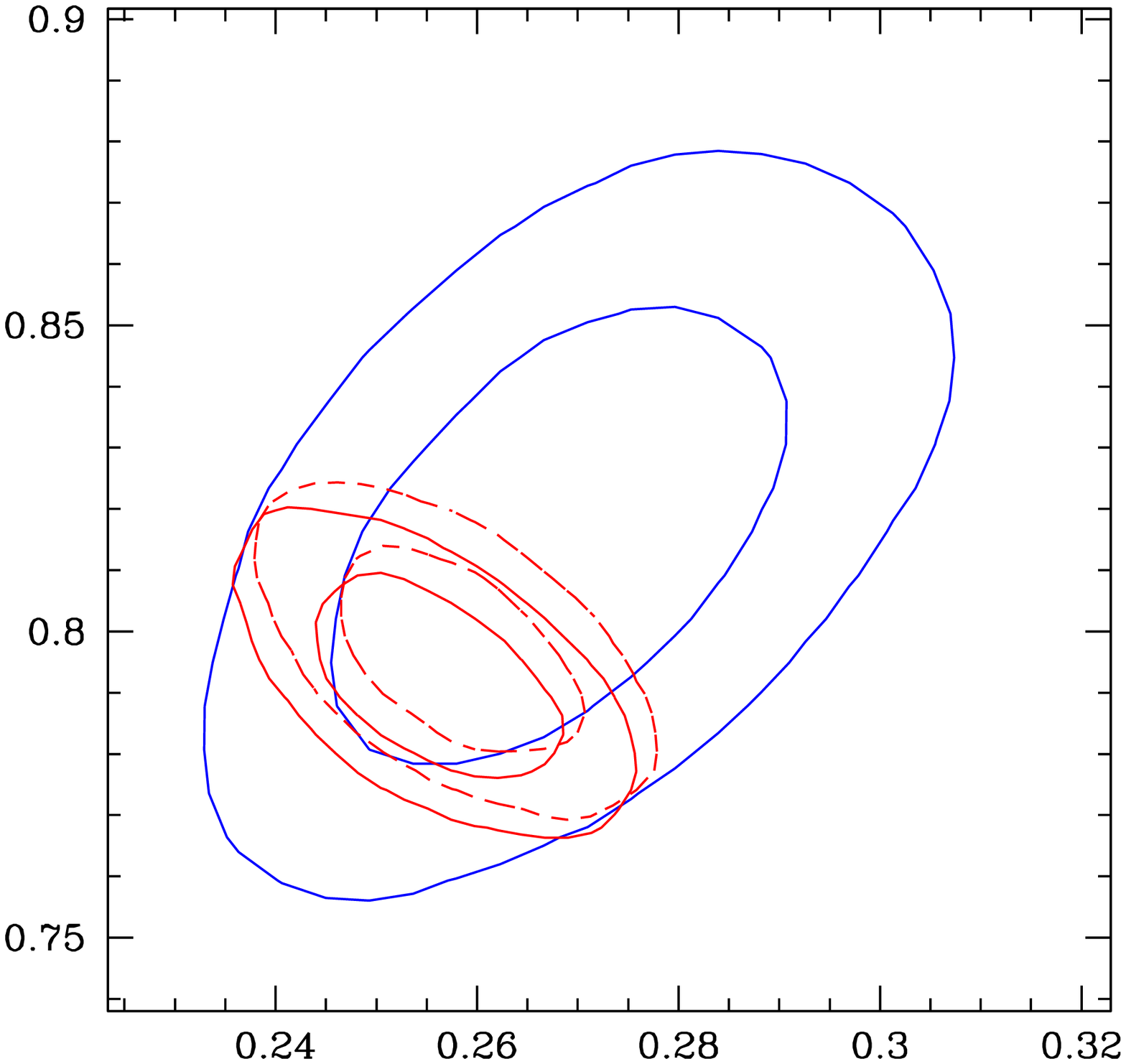}{$\Omega_m$}{$\sigma_8$}
  \end{minipage}
  \caption{Constraints for $\Omega_m$ and $\sigma_8$ in
    \emph{$\Lambda$CDM} model. Contours of larger size --- from
    \emph{WMAP7+BAO+$H_0$} dataset, of smaller size --- from
    \emph{CL+WMAP7+BAO+$H_0$} dataset. Dashed lines show contours for
    $\delta M/M\approx 0.09$ systematical shift in cluster mass
    measurements (left), and also for the departure from self-similar
    evolution, $\delta M/M \approx 0.05$ for distant clusters
    (right).}
  \label{fig:oms8}
\end{figure}

This systematical uncertainty of cluster mass measurements gives the
uncertainty $\delta\sigma_8 \approx 0.02$ for fixed $\Omega_m$
\citep{av09a}. When the other cosmological data are taken in account,
systematical underestimation of cluster masses for $\delta M/M \approx
0.09$ leads to the shift of confidence contours, as it is shown in the
left panel of Fig.~\ref{fig:oms8}. In this Figure, as an example, we
show the constraints in $\Omega_m$ -- $\sigma_8$ plane, in the model
of flat Universe with cosmological constant (\emph{$\Lambda$CDM}),
obtained using \emph{CL+WMAP7+BAO+$H_0$} cosmological data set. In the
left part of this Figure the confidence contours for $\delta
M/M\approx 0.09$ systematical shift in cluster mass measurements are
shown with dashed lines. Since the likelihood maximum is shifted not
along the line of constant $\Omega_m$, the systematic error for
$\sigma_8$ parameter turns out to be substantially smaller,
$\delta\sigma_8 \approx 0.011$ in this case. Therefore, systematic
errors for cosmological parameters may be significantly reduced when
the additional cosmological data are taken in account.

In addition to the error of the mass measurement calibration for
nearby clusters, a significant part of total systematical error comes
from the uncertainties in their cosmological evolution, which lead to
the difference in mass scale calibrations for nearby and distant
clusters. The systematic uncertainty due to the departure from
self-similar evolution can be estimated as $\delta M/M \approx 0.05$
for distant clusters at $z\approx0.6$ \citep[see details
in][]{av09a}. This difference in mass scales corresponds to the shift
of confidence contours in $\Omega_m$--$\sigma_8$ plane shown in right
panel of Fig.~\ref{fig:oms8}. Quadratic sum of to systematic errors
estimated for these two main uncertainties in cluster mass
measurements is given below as an estimate of total systematical error
for cosmological parameters constraints.

In correspondent parts of this paper the systematic uncertainties of
the data on observations of supernovae Type Ia, \emph{UNION2}
compilation \citep{amanullah10}, are also considered.  Fore these data
the total, systematical and statistical errors were calculated using
correspondent covariance matrix. Separate estimate of systematic
error for these data was obtained by quadratic subtraction of
statistical error from total one.

\begin{table*}
  \renewcommand{\arraystretch}{1.2}
  \centering
  \caption{Cosmological parameters constraints}
  \label{tab:pars}
  \medskip
  \small
  \begin{tabular}{lllll}
    \hline
    \hline
    Model & Data & Parameter & Value$^{*}$ & 
    Systematical error\\
    \hline
    \emph{$\Lambda$CDM} & 
    \emph{CL+WMAP7+BAO+$H_0$}
    & $\Omega_m h^2$ & $0.1311\pm 0.0016$ & $\pm 0.0018$\\
    \dots & \dots & $z_{eq}$   & $3155\pm38$ & $\pm 44$\\
    \dots & \dots & $\sigma_8$ & $0.792\pm0.010$ & $\pm 0.013$\\
    \dots & \dots & $\Omega_m$ & $0.2558\pm0.0077$ & $\pm 0.0063$\\
    \dots & \dots & $H_0$      & $71.6\pm 1.0$ km~s$^{-1}$~Mpc$^{-1}$ &  $\pm 0.4$ km~s$^{-1}$~Mpc$^{-1}$ \\
    \dots & \dots & $\Omega_b$ & $0.04386\pm0.00086$ & $\pm 0.00064$\\
%    & & $t_0$      & $13.75\pm0.10$ Глет \\
    \hline
    \emph{$\Lambda$CDM} & 
    \emph{CL+WMAP7+BAO+$H_0$+SPT}
    & $\Omega_m h^2$ & $0.1310\pm 0.0014$ & $\pm 0.0016$\\ 
    \dots & \dots & $z_{eq}$   & $3154\pm33$ & $\pm 39$\\  
    \dots & \dots & $\sigma_8$ & $0.7921\pm0.0089$ & $\pm 0.0115$\\
    \dots & \dots & $\Omega_m$ & $0.2560\pm0.0067$ & $\pm 0.0072$\\
    \dots & \dots & $H_0$      & $71.55\pm 0.77$ km~s$^{-1}$~Mpc$^{-1}$ &  $\pm 0.56$ km~s$^{-1}$~Mpc$^{-1}$ \\
    \dots & \dots & $\Omega_b$ & $0.04349\pm0.00075$ & $\pm 0.00071$\\
    
    % \multirow{5}{*}{\emph{$\Lambda$CDM}} & 
    % \multirow{5}{*}{\emph{CL+WMAP7+BAO+$H_0$+SPT}}
    % & $\Omega_m h^2$ & $0.1324\pm 0.0014$ & $\pm 0.0010$\\ 
    % & & $z_{eq}$   & $3187\pm33$ & $\pm 24$\\ 
    % & & $\sigma_8$ & $0.7958\pm0.0092$ & $\pm 0.0068$\\
    % & & $\Omega_m$ & $0.2523\pm0.0067$ & $\pm 0.0043$\\ 
    % & & $H_0$      & $72.46\pm 0.81$ км~с$^{-1}$~Мпк$^{-1}$ &  $\pm 0.34$ км~с$^{-1}$~Мпк$^{-1}$ \\
    % & & $\Omega_b$ & $0.04380\pm0.00073$ & $\pm 0.00044$\\%0.0438  0.00073   0.000436
    \hline

%--- omegak:

 %   \emph{$\Lambda$CDM}+$\Omega_k$ &
 %    \emph{CL+WMAP7+BAO+$H_0$} & $\Omega_k$ & $-0.0022\pm0.0044$ & \\
 %   \dots & \dots & $\Omega_\Lambda$ & $0.7439\pm0.0079$ & \\

    \emph{$\Lambda$CDM}+$\Omega_k$ & \emph{CL+WMAP7+BAO+$H_0$+SPT} & $\Omega_k$ & $-0.0018\pm0.0042$ & \\
    \dots & \dots & $\Omega_\Lambda$ & $0.7430\pm0.0072$ & \\
%    \emph{$\Lambda$CDM}+$Y_{p}$
%    & \dots
%    & $Y_{p}$ & $0.279\pm0.028$ & $\pm0.008$\\
    
    \hline
    \emph{$\Lambda$CDM}+$m_\nu$ & 
%    \emph{CL+WMAP7+$H_0$} & $\Sigma m_\nu$ & $<0.29$ eV \\
%    \dots & 
    \emph{CL+WMAP7+$H_0$+BAO+SPT} & $\Sigma m_\nu$ & $<0.32$ eV \\
    \dots & 
    \emph{CL+WMAP7+$H_0$+BAO+SPT+SN} & $\Sigma m_\nu$ & $<0.28$ eV \\

%--- neff:
    \emph{$\Lambda$CDM}+$\neff$ & 
%    \emph{CL+WMAP7+BAO+$H_0$} & $\neff$ & $<4.07$ &\\
%    \dots & 
    \emph{CL+WMAP7+BAO+$H_0$+SPT} & $\neff$ & $<3.74$\\
    \dots & \emph{CL+WMAP7+BAO+$H_0$+SPT+SN} & $\neff$ & $<3.70$\\

%--- mnu+neff:
    \emph{$\Lambda$CDM}+$m_\nu$+$\neff$ & 
%    \emph{CL+WMAP7+BAO+$H_0$} & $\Sigma m_\nu$ & $<0.75$ eV \\
%    \dots & \dots & $\neff$ & $<5.19$ & \\
%    \dots & 
    \emph{CL+WMAP7+BAO+$H_0$+SPT} & $\Sigma m_\nu$ & $<0.72$ eV\\
    \dots & \dots & $\Sigma m_\nu$ & $0.36\pm 0.16^{**}$ eV & $\pm 0.08$ eV\\
     \dots & \dots & $\neff$ & $<4.62$  & \\
     
%    \emph{$\Lambda$CDM}+$r$
%    & \multirow{2}{*}{\emph{CL+WMAP7}}
%    & $r$ & $<0.26$ (д.\ у.\ 95\%)\\
%    \emph{$\Lambda$CDM}+$dn_s/d \ln(k)$
%    &
%    & $dn_s/d \ln(k)$ & $-0.015\pm0.021$ \\
     \hline
    \emph{WCDM} & \emph{CL+WMAP7+BAO+$H_0$}
    & $w$ & $-1.027 \pm 0.069$ & $\pm 0.028$ \\
    \dots & \emph{CL+WMAP7+BAO+$H_0$+SN}
    & $w$ & $-0.990 \pm 0.034$ & $\pm 0.041^{***}$\\ 
    \dots & \emph{CL+WMAP7+BAO+$H_0$+SPT}
    & $w$ & $-1.013 \pm 0.066$ & $\pm 0.029$ \\
    \dots & \emph{CL+WMAP7+BAO+$H_0$+SPT+SN}
    & $w$ & $-0.982 \pm 0.032$ & $\pm 0.038^{***}$\\%sqrt(0.04569**2-0.03212**2+0.019581**2)
    \emph{WCDM}+$\Omega_k$ & \emph{CL+WMAP7+BAO+$H_0$}
    & $w$ & $-1.12\pm 0.13$ & \\ 
    \dots & \emph{CL+WMAP7+BAO+$H_0$+SN}
    & $w$ & $-0.991\pm 0.039$ & \\ 
    \emph{WACDM} & \emph{CL+WMAP7+BAO+$H_0$+SN}
    & $w_0$ & $-1.13\pm  0.11$\\
    \dots & \dots 
    & $w_a$ & $0.47\pm0.36$\\
    \hline
  \end{tabular}
  
  \medskip
  \begin{minipage}{0.95\linewidth}
    $^{*}$ --- all intervals are given at 68\% confidence level, all
    upper limits --- at 95\% confidence level;
    
    $^{**}$ --- interval is non-gaussian, for significance of non-zero
    value see text;
    
    $^{***}$ --- including systematic uncertainty of SNe Ia data;
  \end{minipage}

\end{table*}

\begin{figure*}
  {\scriptsize \hskip 8mm \emph{\textcolor{red}{CL+} \textcolor{darkblue}{WMAP7}}
    \hskip 2.83cm \emph{\textcolor{red}{CL+} \textcolor{darkblue}{WMAP7+$H_0$}}
    \hskip 2.25cm \emph{\textcolor{red}{CL+} \textcolor{darkblue}{WMAP7+$H_0$+BAO}}
    \hskip 1.63cm \emph{\textcolor{red}{CL+} \textcolor{darkblue}{WMAP7+$H_0$+BAO+SPT}}
  }

  \centering
  \begin{minipage}{0.24\linewidth}
    \smfigure{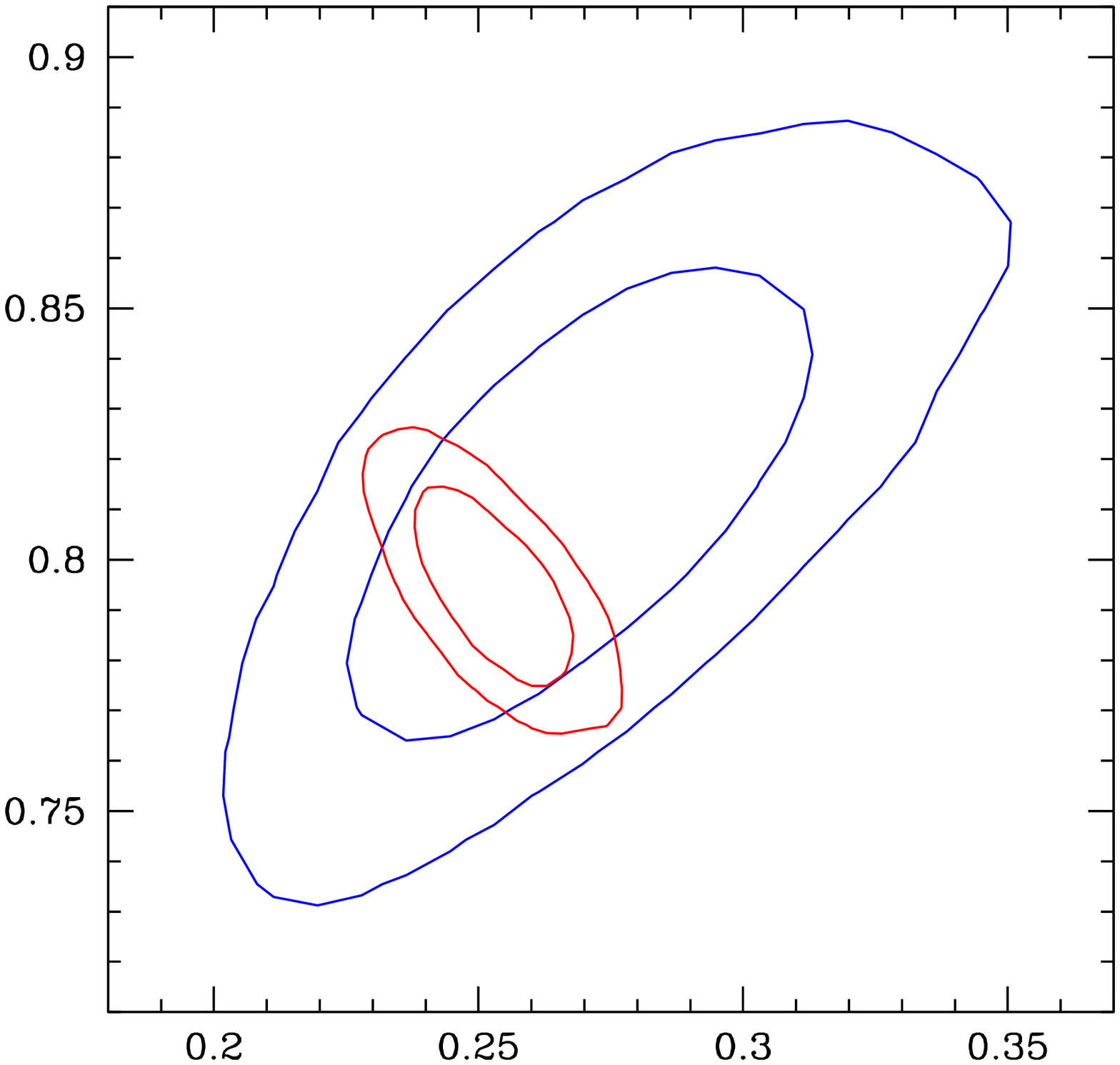}{$\Omega_m$}{$\sigma_8$}
  \end{minipage}
  \begin{minipage}{0.24\linewidth}
    \smfigure{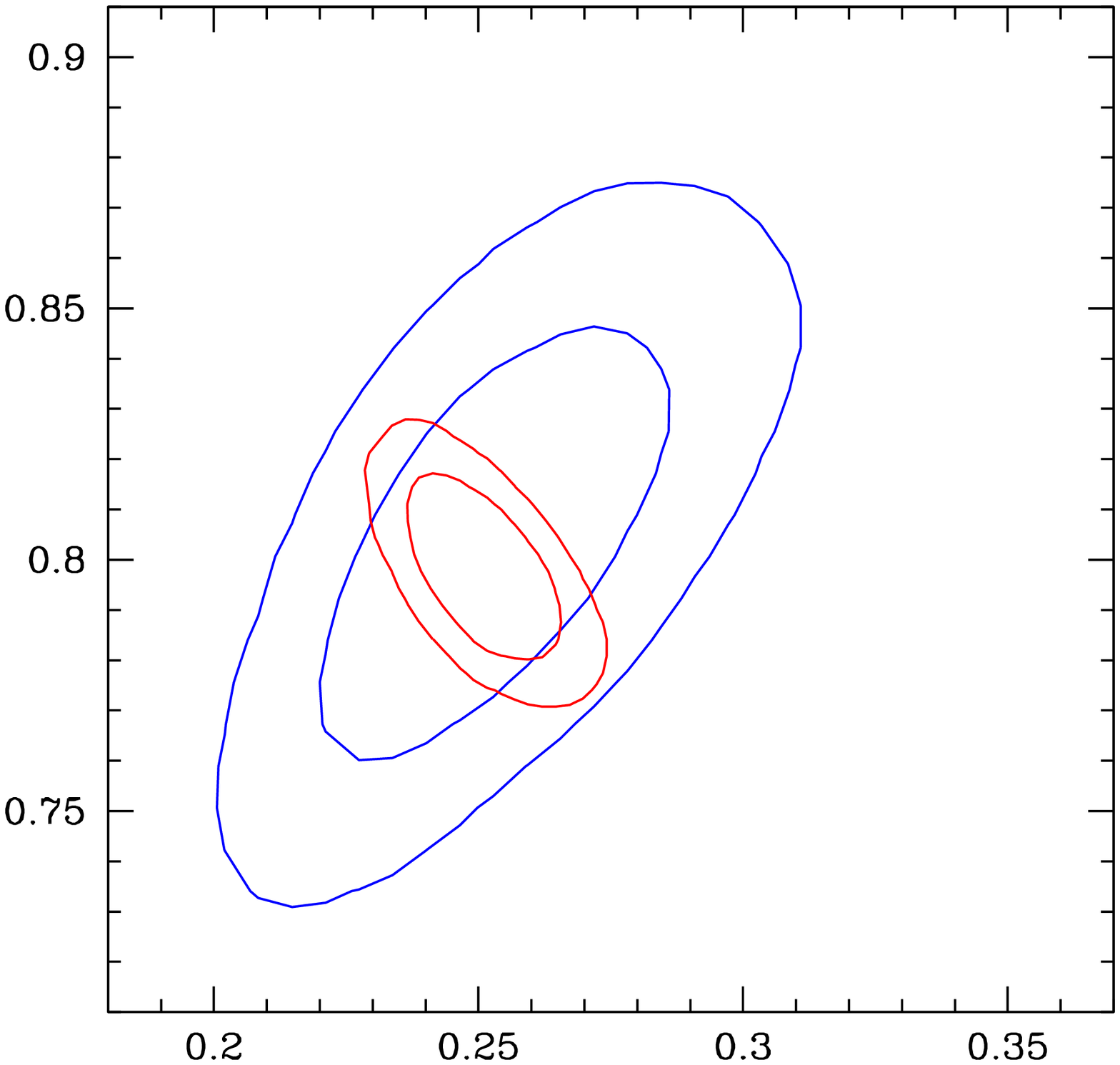}{$\Omega_m$}{$\sigma_8$}
  \end{minipage}
  \begin{minipage}{0.24\linewidth}
    \smfigure{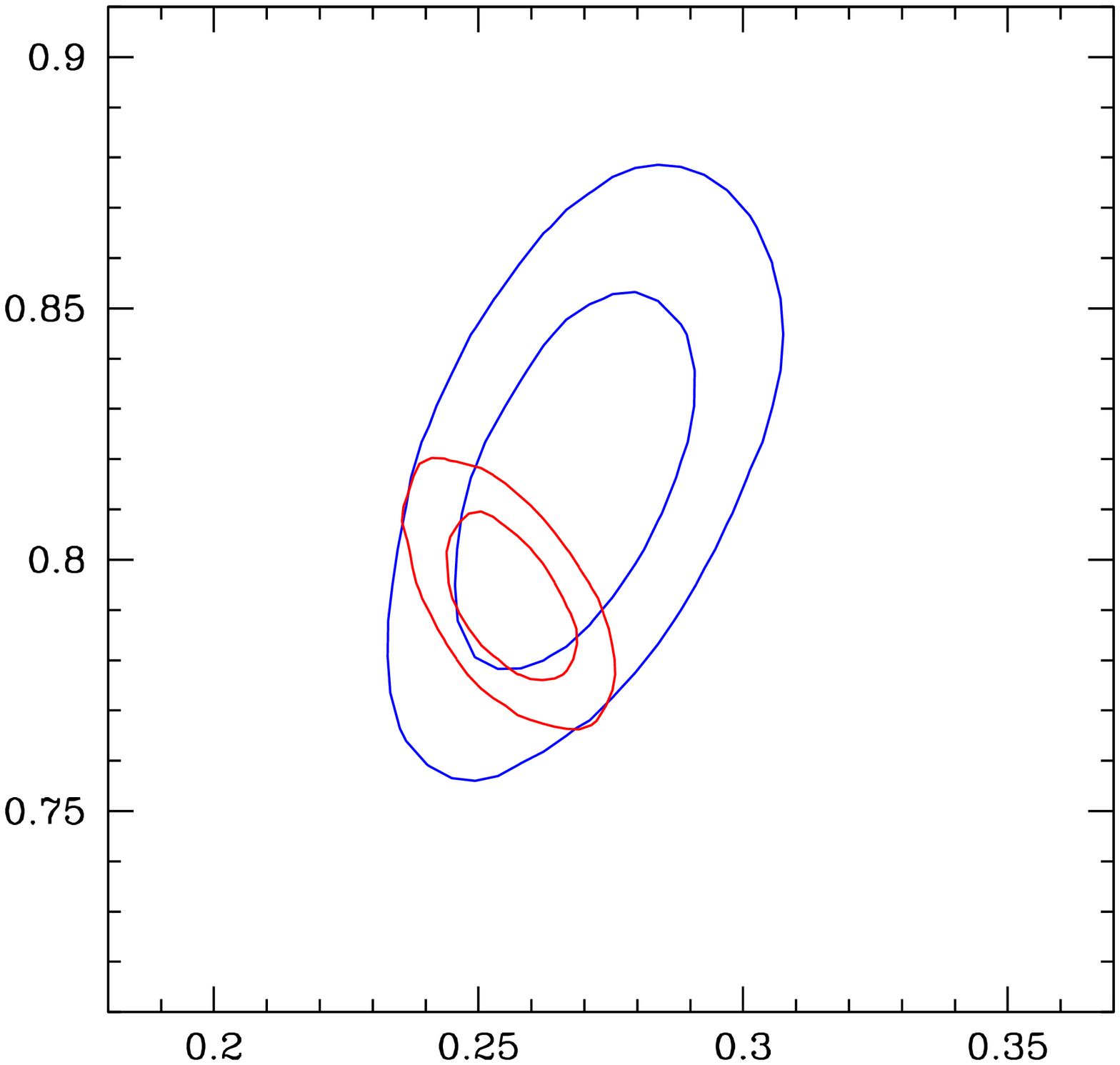}{$\Omega_m$}{$\sigma_8$}
  \end{minipage}
  \begin{minipage}{0.24\linewidth}
    \smfigure{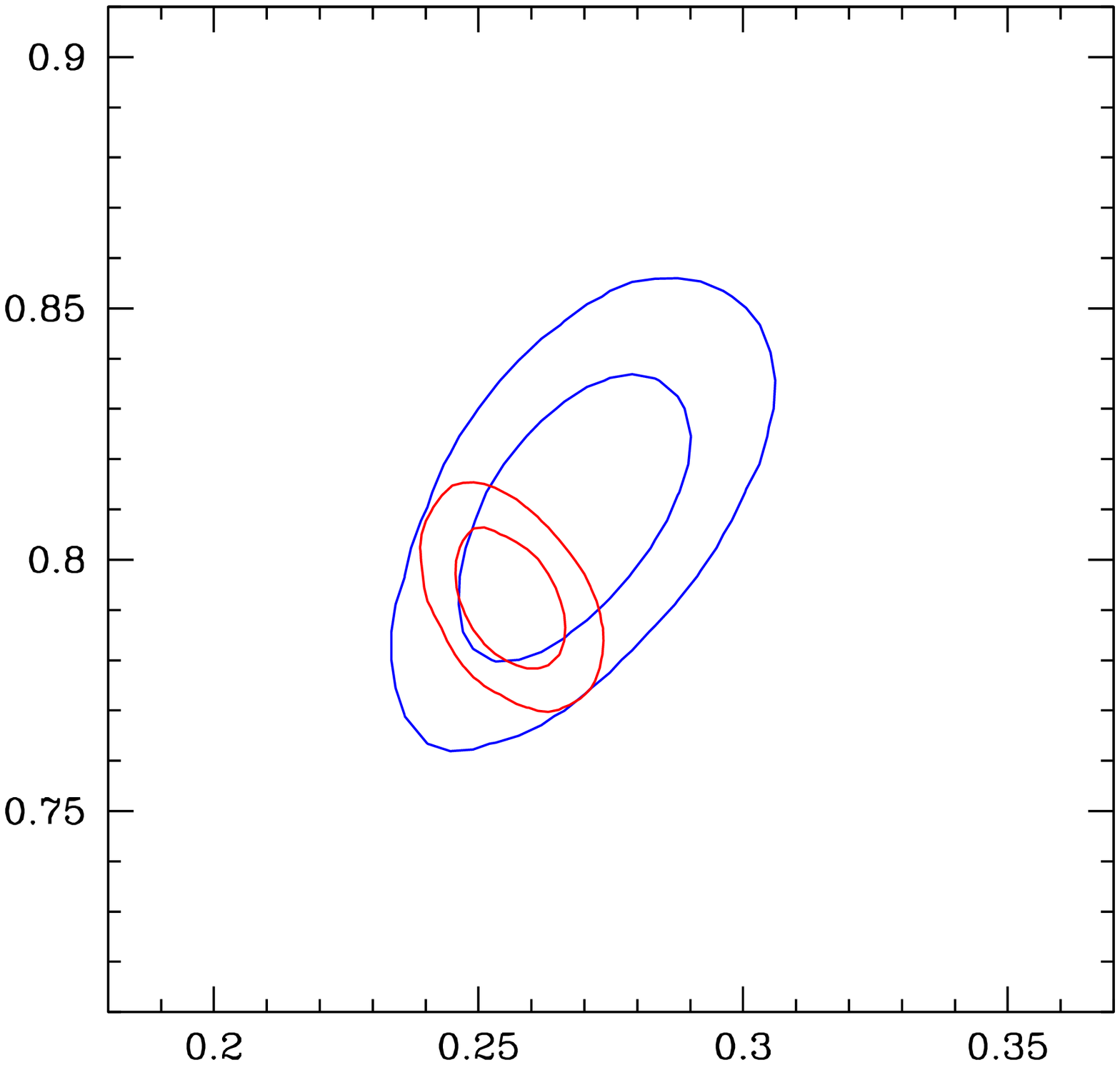}{$\Omega_m$}{$\sigma_8$}
  \end{minipage}
  
  \caption{Constraints on $\Omega_m$ and $\sigma_8$ in
    \emph{$\Lambda$CDM} model, obtained using different
    datasets. Larger contours, from left to the right: \emph{WMAP7},
    \emph{WMAP7+$H_0$}, \emph{WMAP7+BAO+$H_0$},
    \emph{WMAP7+BAO+$H_0$+SPT}. Smaller contours --- using the same
    data, with the data on galaxy cluster mass function added.}

  \label{fig:oms8comp}
\end{figure*}

\begin{figure}
  \centering
  \begin{minipage}{0.48\linewidth}
    \smfigure{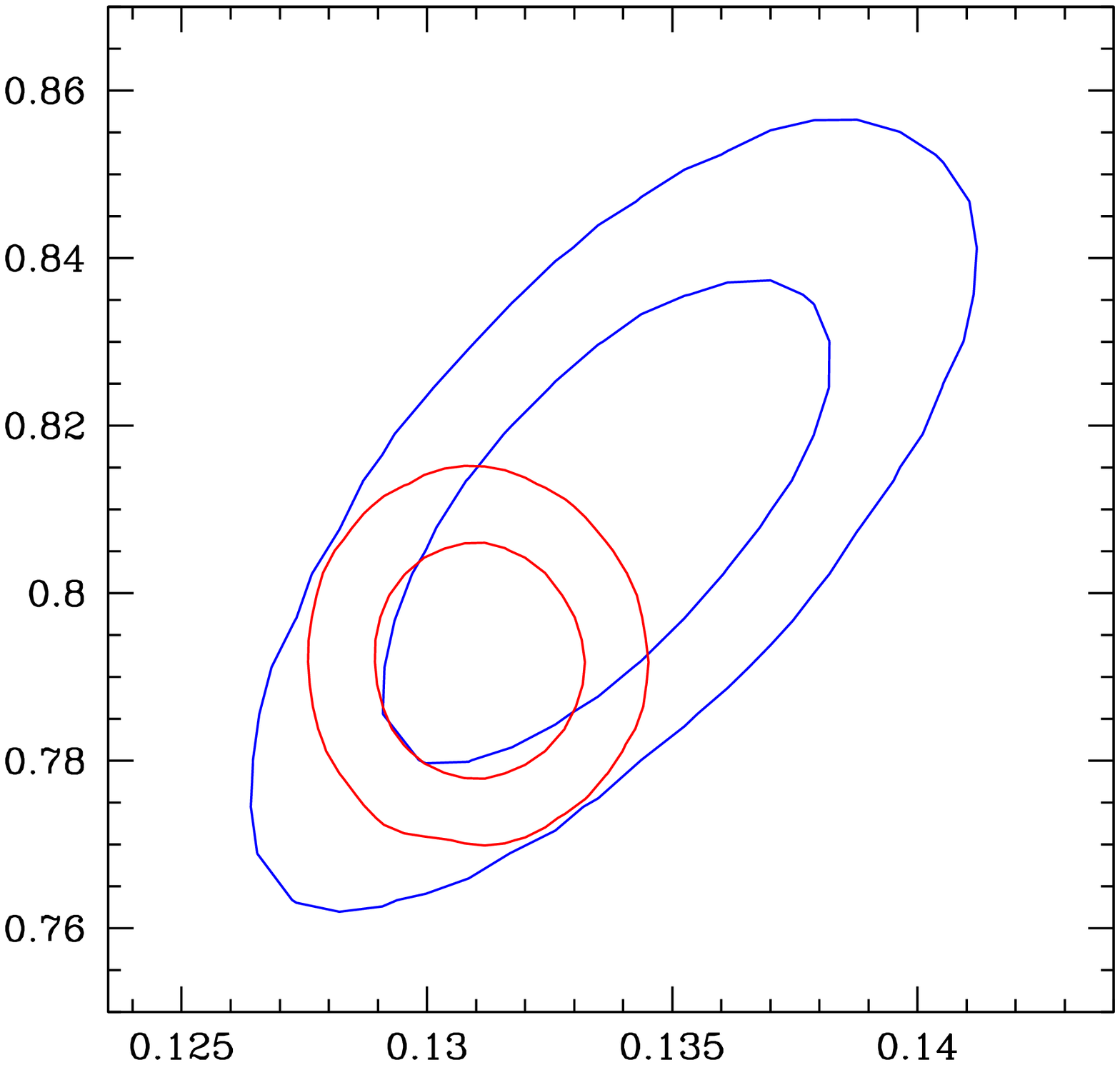}{$\Omega_m h^2$}{$\sigma_8$}
  \end{minipage}
  \begin{minipage}{0.48\linewidth}
    \smfigure{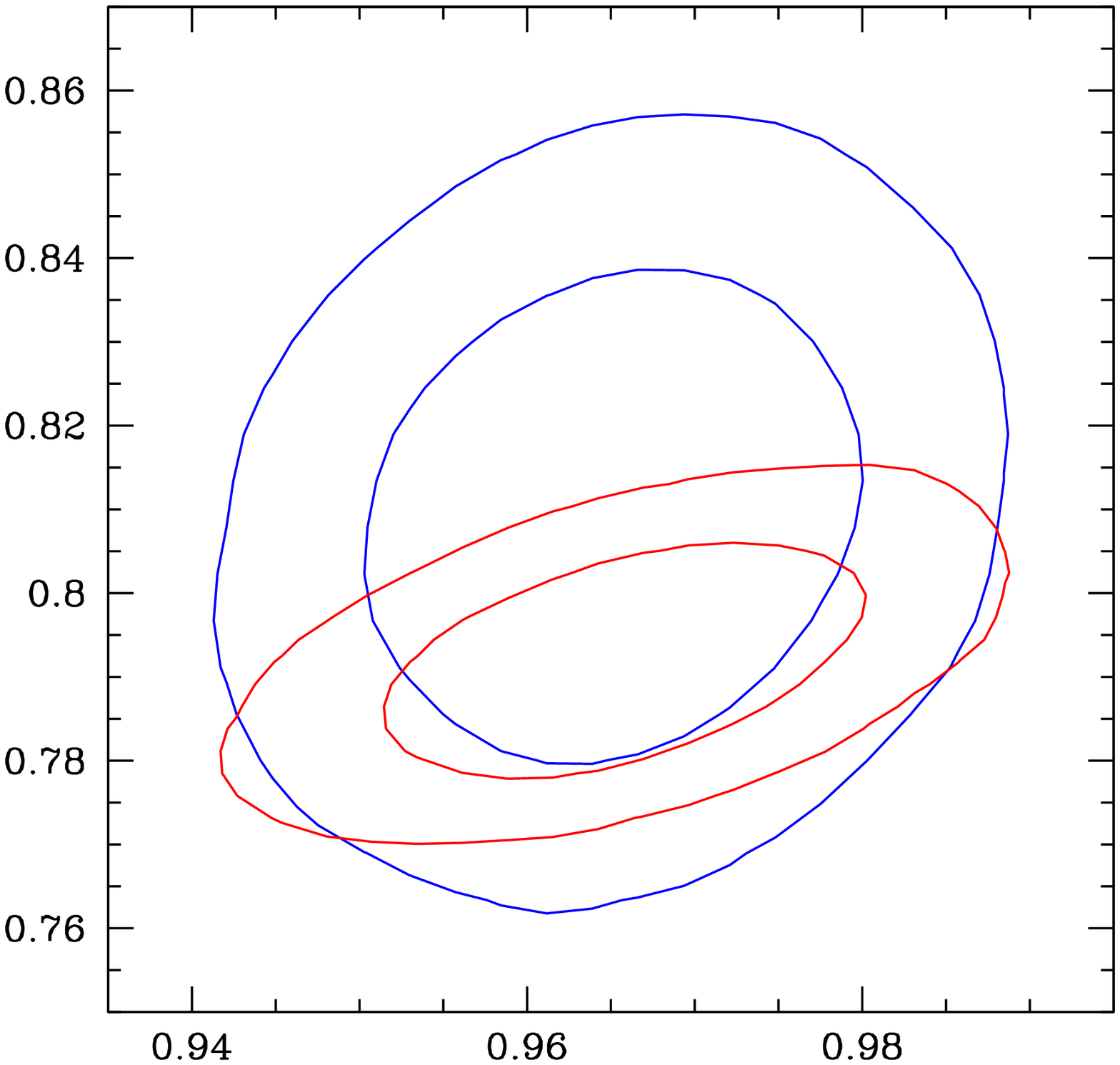}{$n_{s}$}{$\sigma_8$}
  \end{minipage}

   \begin{minipage}{0.48\linewidth}
    \smfigure{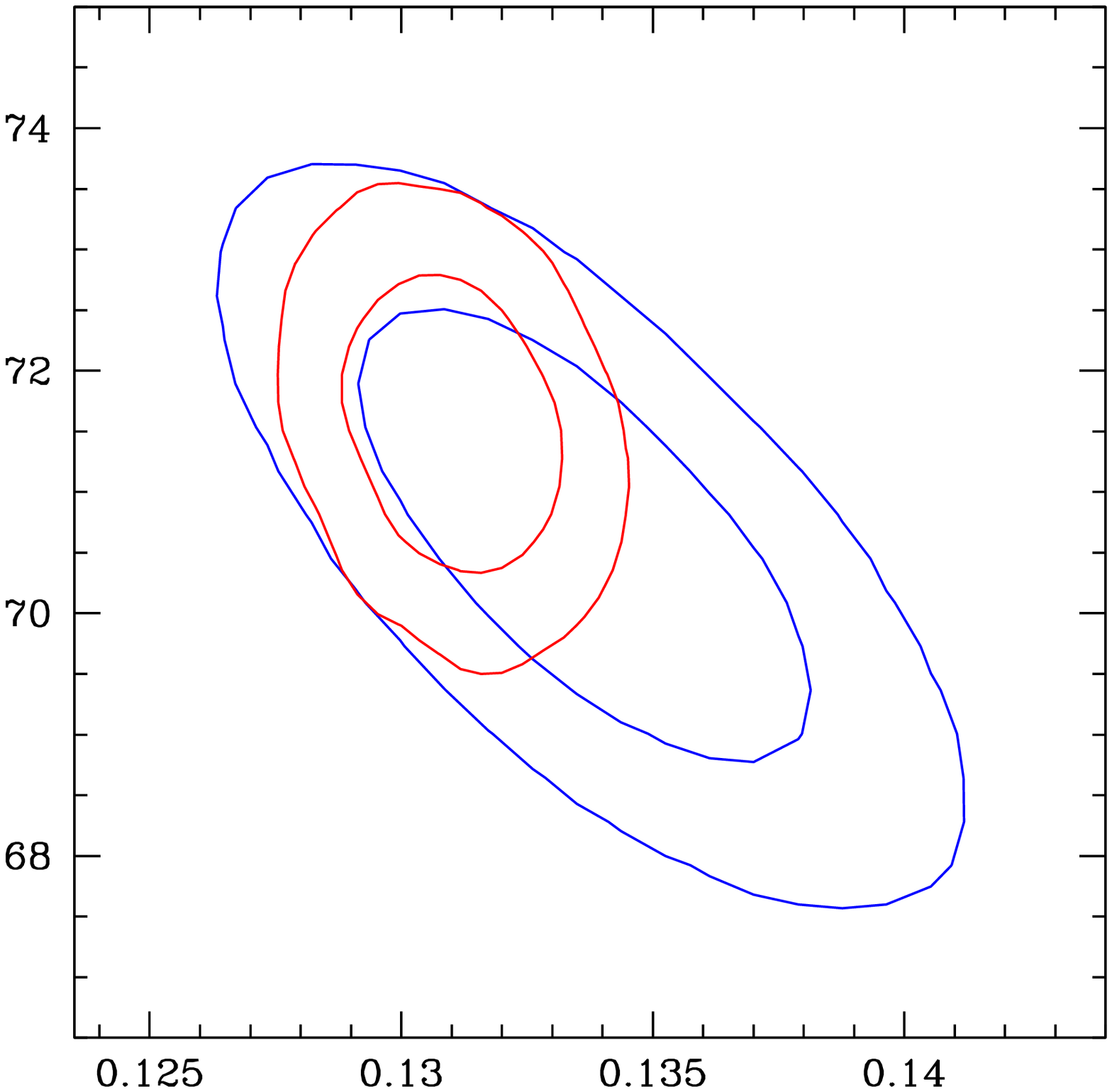}{$\Omega_m h^2$}{$H_0$}
  \end{minipage}
  \begin{minipage}{0.48\linewidth}
    \smfigure{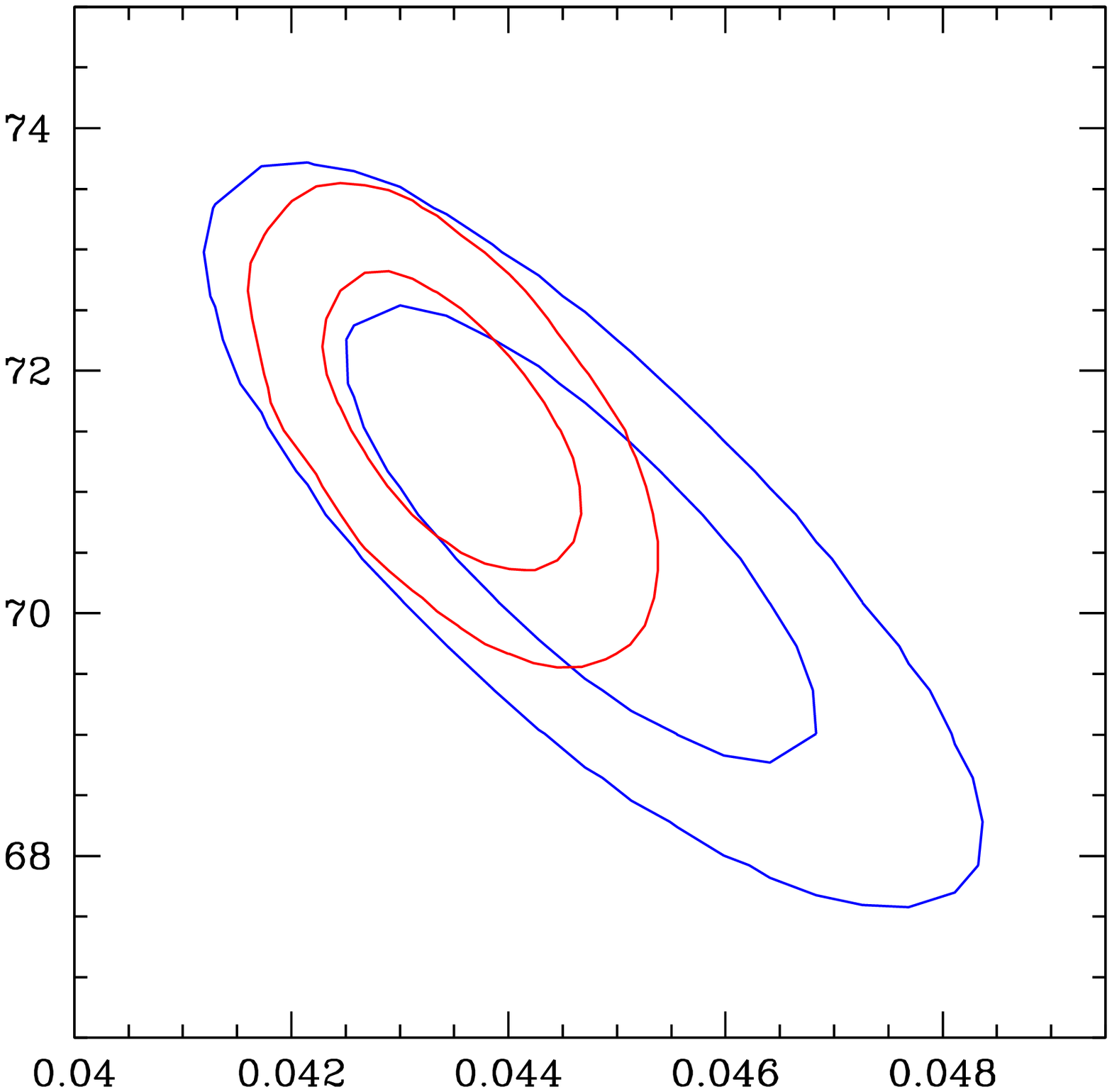}{$\Omega_b$}{$H_0$}
  \end{minipage}

  \caption{Constraints on $\Omega_m h^2$, $\sigma_8$, $H_0$,
    $\Omega_b$ and $n_{s}$ in \emph{$\Lambda$CDM} model. Contours of
    larger size --- from \emph{WMAP7+BAO+$H_0$+SPT} dataset, of
    smaller size --- from \emph{CL+WMAP7+BAO+$H_0$+SPT} dataset.}
  \label{fig:omh2s8}
\end{figure}

\section{Constraints for \emph{$\Lambda$CDM} model}
\label{sec:lcdm}

In the upper part of Table~\ref{tab:pars} the cosmological parameters
constraints for flat Universe with cosmological constant
(\emph{$\Lambda$CDM} model) are presented. The nature of main
constraints is as follows. The constraints from galaxy cluster mass
function are based mainly on its normalization and have their main
effect on $\Omega_m$ and $\sigma_8$ --- the constraints turns out to
be degenerate, these two parameters are related as $\sigma_8 =
0.813(\Omega_M /0.25)^{-0.47} \pm 0.011$ \citep{av09b}. The relation
between $\sigma_8$ and CMB anisotropy amplitude depends mostly on
linear perturbation growth between decoupling epoch and $z=0$
\citep[see, e.g., approximate formula from][]{hujain04}, which in turn
strongly depends on $\Omega_m$. Therefore, the data on CMB anisotropy
amplitude allow to break the above $\Omega_m$--$\sigma_8$ degeneracy
and these data analyzed jointly provide most accurate measurement of
both $\sigma_8$ and $\Omega_m$.

The relative power of various datasets to constrain $\Omega_m$ and
$\sigma_8$ parameters is shown in Fig.~\ref{fig:oms8comp}. One can see
that the main power of constraints in \emph{$\Lambda$CDM} model indeed
originate from the joint analysis of galaxy cluster mass function and
CMB data. The other cosmological data give only smaller improvement to
the joint constraints on $\sigma_8$ and $\Omega_m$.

Parameters $\sigma_8$ and $\Omega_m h^2$ appear to be best constrained
from joint analysis of all cosmological data considered in our work,
which is shown in Fig.~\ref{fig:omh2s8}. The projection of these
constraints on other parameters significantly improve their
measurements in some cases. For example, the constraints are improved
for parameters $H_0$ and $\Omega_b$. However, the data on cluster mass
function do not improve significantly, for example, the constraints on
the combination $\Omega_b h^2$, or on spectral index of density
perturbations $n_s$.

Note, that in considered cosmological model it is suggested that there
are no other relativistic particles at equipartition in addition to
photons and three known neutrino species and the relativistic energy
density is known exactly. Therefore, in this model the constraint on
$\Omega_m h^2$ is equivalent to the constrain on equipartition
redshift $z_{eq}$ (in the left panel of Fig.~\ref{fig:omh2s8} the axis
$\Omega_m h^2$ may be changed to $z_{eq}$, see also
Table~\ref{tab:pars}). If the number of relativistic particles is
considered as free parameter, relativistic energy density is no longer
exactly defined quantity and it may be measured from cosmological
data. In this case, the data on galaxy cluster mass function allow to
obtain significant constraints on the number of relativistic species
(see below).

\begin{figure}
  \centering
  \begin{minipage}{0.65\linewidth}
    \smfigure{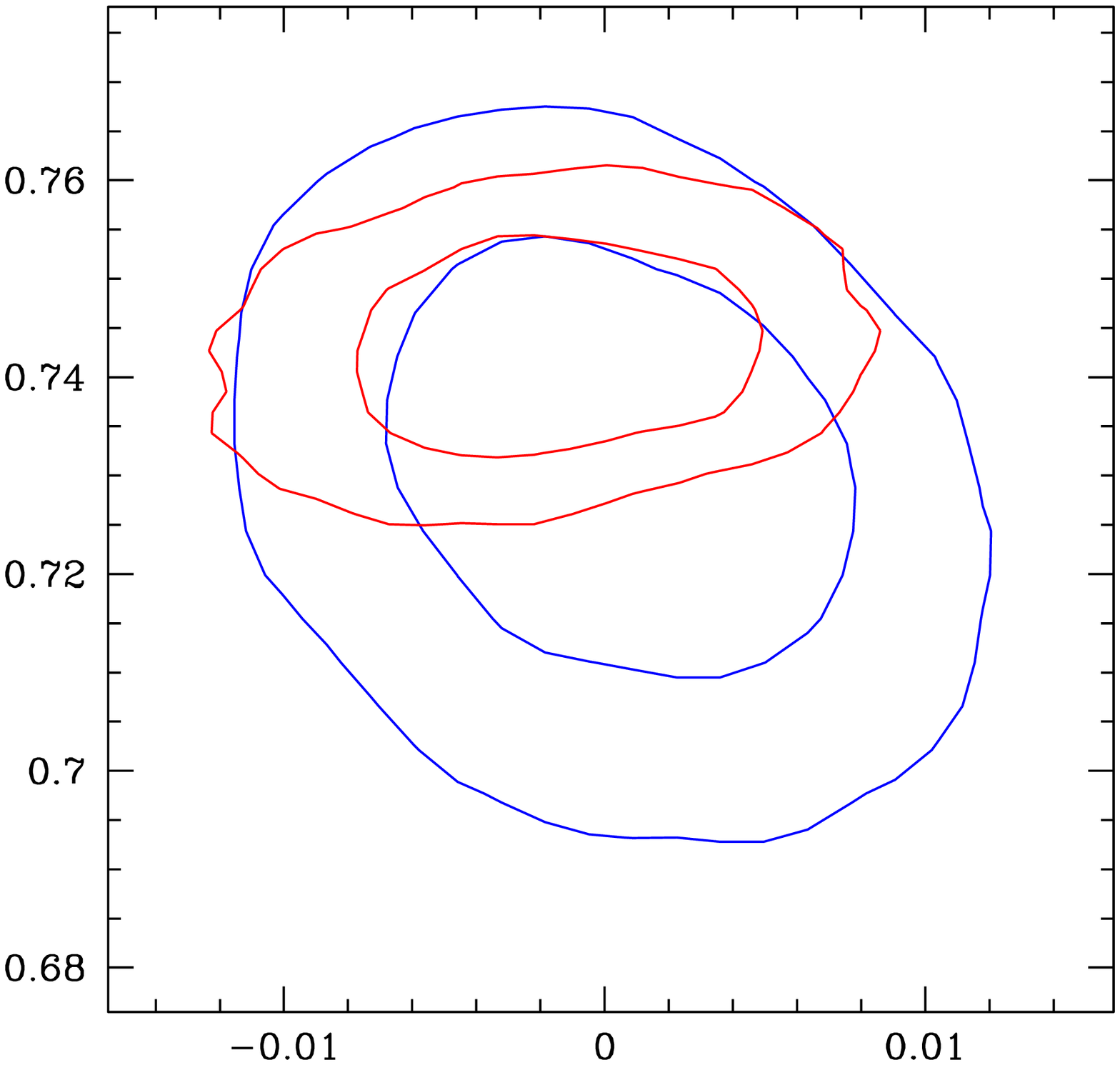}{$\Omega_k$}{$\Omega_\Lambda$}
  \end{minipage}
  \caption{Constraint on spacial curvature in Universe, $\Omega_k$, in
    \emph{$\Lambda$CDM} model with $\Omega_k\ne 0$. Contours are the
    same as in Fig.~\ref{fig:omh2s8}.}
  \label{fig:ok}
\end{figure}

\section{Extensions of \emph{$\Lambda$CDM} model}
\label{sec:lcdmdev}

\subsection{Spacial curvature}
\label{sec:curvature}

Constraints on spacial curvature in \emph{$\Lambda$CDM} model with
$\Omega_k\ne 0$, and also on the cosmological constant density are
shown in Fig.~\ref{fig:ok} (see also Table~\ref{tab:pars}). The
constraint on $\Omega_k$ is not improved significantly, there is only
approximately 20\% improvement, as compared to the case where cluster
data are not taken in account (see Fig.~\ref{fig:ok}). We note, that
in this model galaxy cluster data allow to better constrain $\Omega_m$
and $\Omega_\Lambda$ separately.

\subsection{Primordial helium abundance}
\label{sec:yhe}

\begin{figure}
  \centering
  \begin{minipage}{0.49\linewidth}
    \smfigure{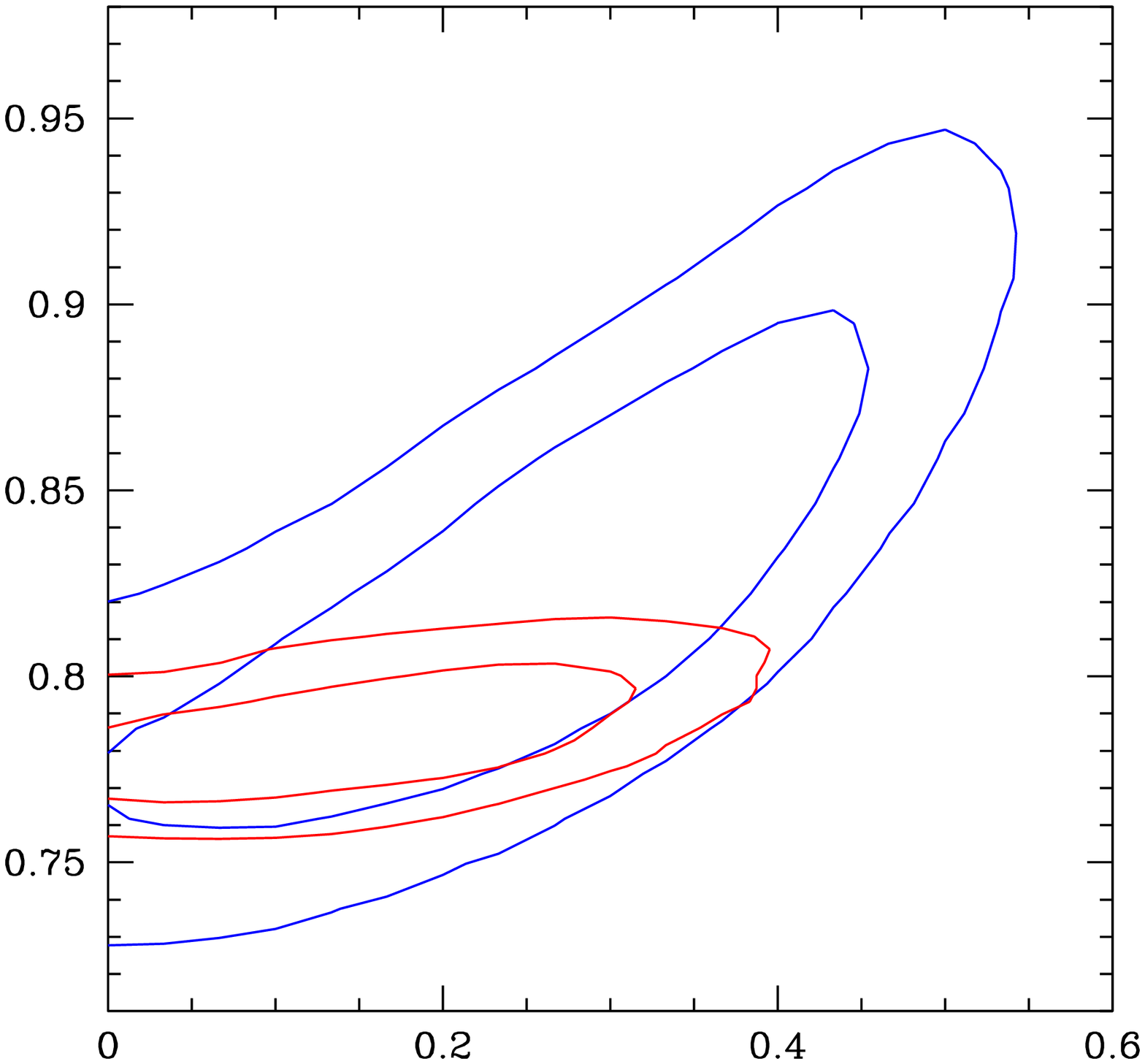}{$Y_{p}$}{$\sigma_8$}
  \end{minipage}
  \begin{minipage}{0.49\linewidth}
    \smfigure{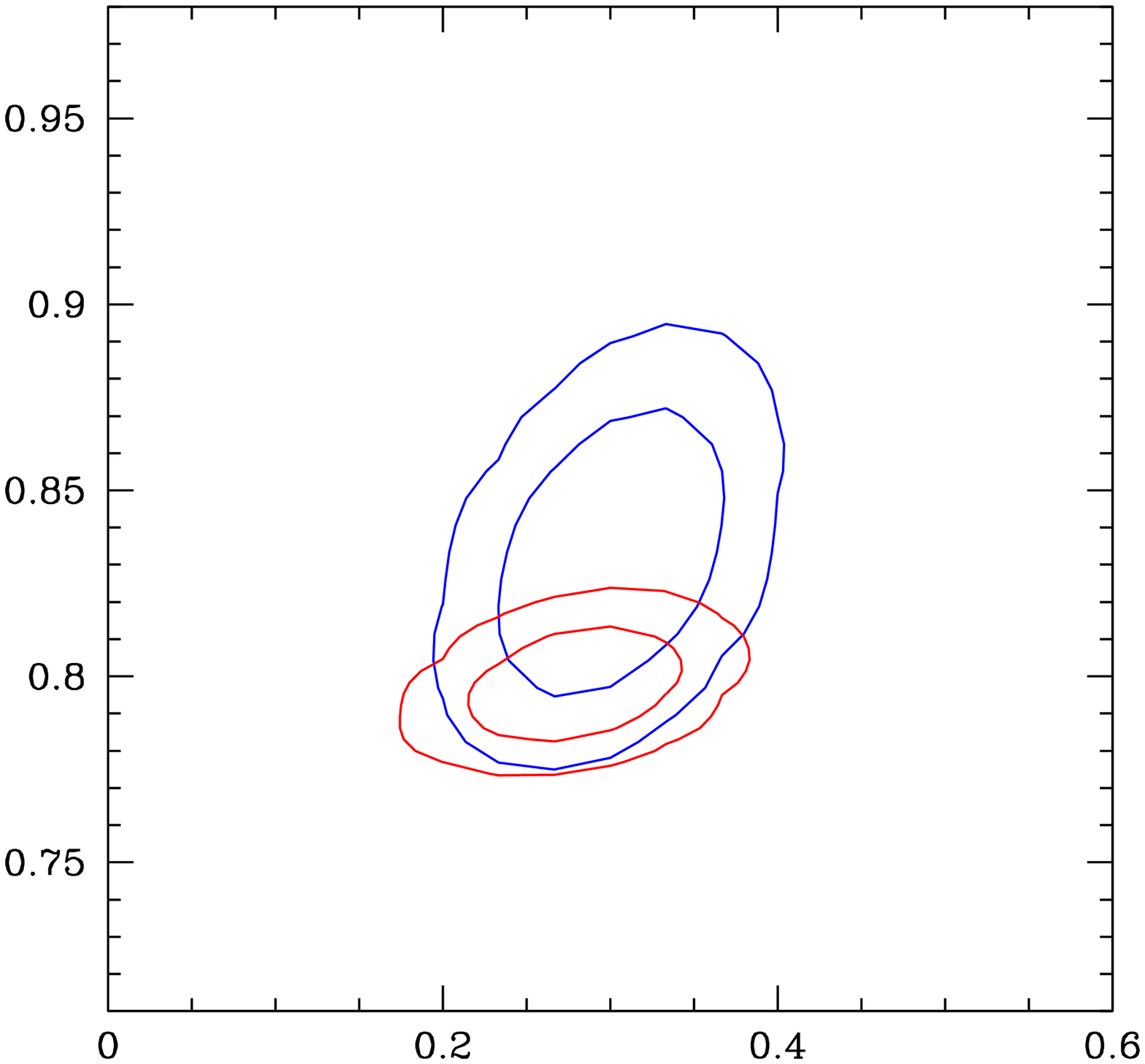}{$Y_{p}$}{$\sigma_8$}
  \end{minipage}
  \caption{Constraints on primordial helium abundance in
    \emph{$\Lambda$CDM} model with free $Y_{p}$. Larger contours ---
    \emph{WMAP7+BAO+$H_0$} dataset (left), and from
    \emph{WMAP7+BAO+$H_0$+SPT} dataset (right), smaller contours
    --- using the same data, with the data on galaxy cluster mass
    function added.}
  \label{fig:yhe}
\end{figure}

If the data on galaxy cluster mass function are added to
\emph{WMAP7+BAO+$H_0$} dataset, the upper limit on primordial helium
abundance, $Y_{p}$, is significantly improved (see Fig.~\ref{fig:yhe},
left panel). However, the data on cluster mass function do not change
significantly the constraints, which are obtained with \emph{SPT} data
taken in account (Fig.~\ref{fig:yhe}, right panel). From
\emph{CL+WMAP7+BAO+$H_0$+SPT} dataset we obtain $Y_{p}=0.279\pm0.028$,
i.e., the measurement of primordial helium abundance is shifted
slightly to its standard value $Y_{p}\approx0.25$, which is obtained
from the theoretical calculations of primordial nucleosynthesis and
from the measurements of $\Omega_b h^2$ with standard value of
effective number of neutrino species $\neff$.

\subsection{Tensor modes and running spectral index}
\label{sec:r}

\begin{figure}
  \centering
  \begin{minipage}{0.49\linewidth}
    \smfigure{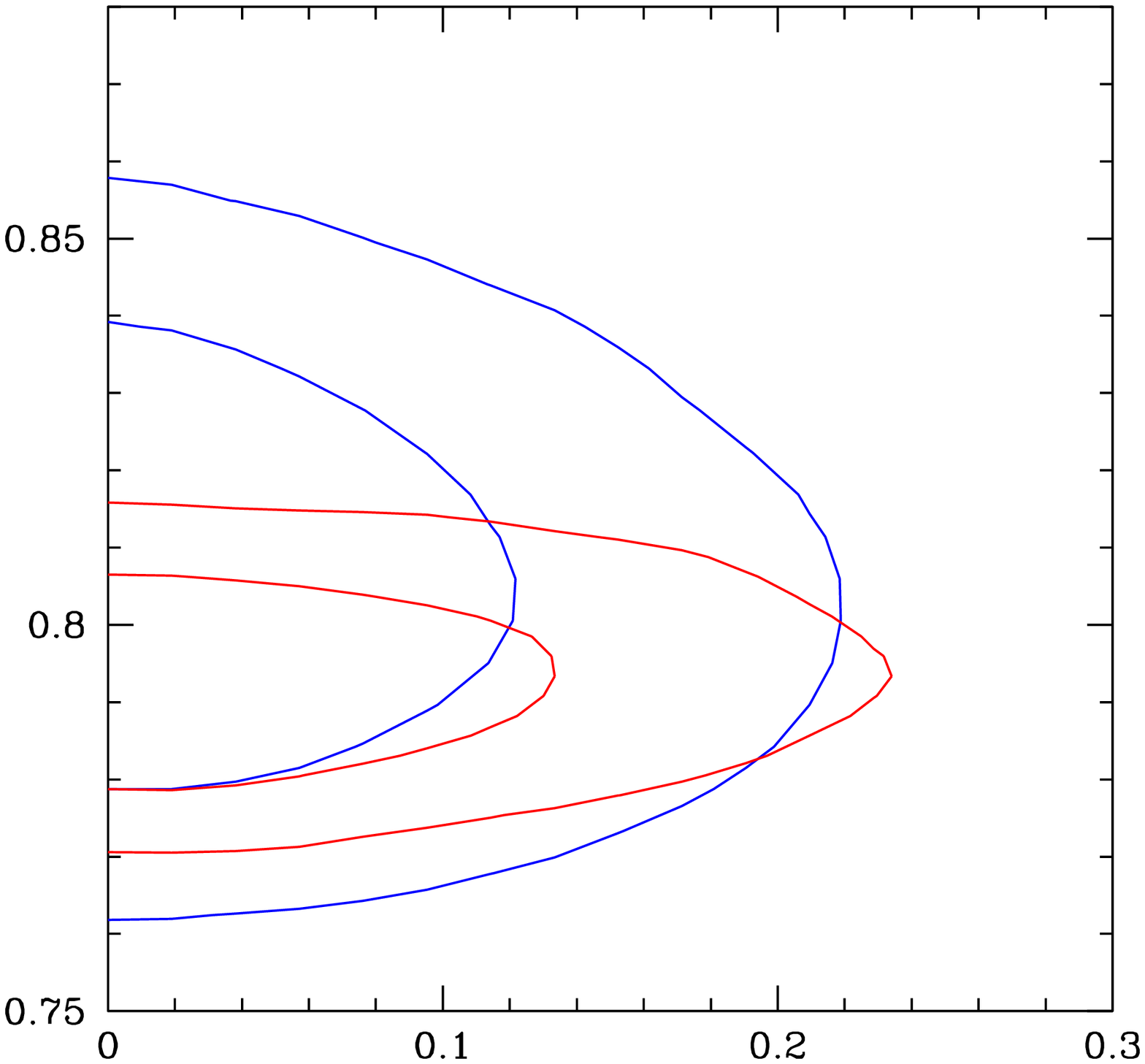}{$r$}{$\sigma_8$}
  \end{minipage}
  \begin{minipage}{0.49\linewidth}
    \smfigure{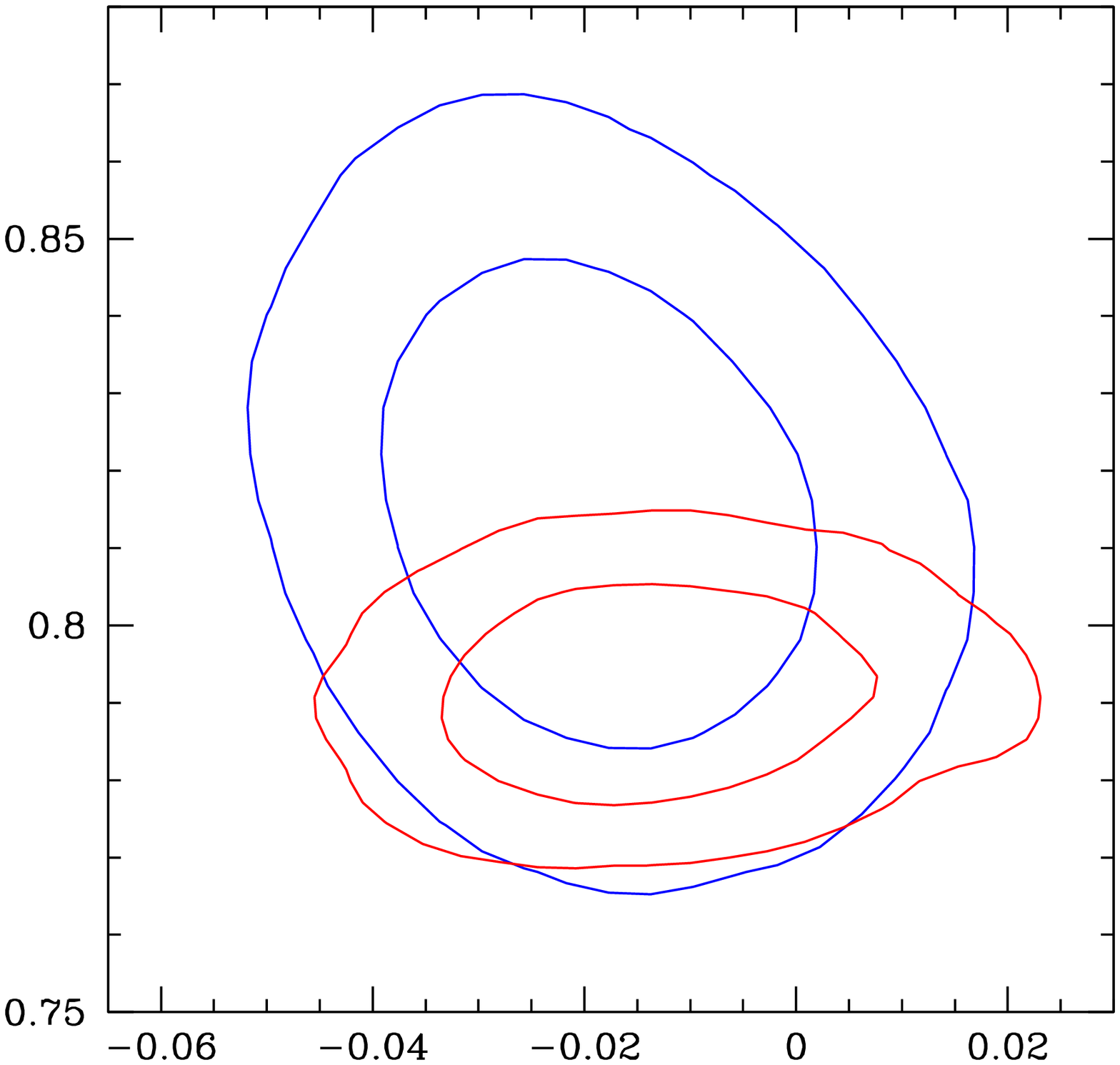}{$d n_s / d \ln k$}{$\sigma_8$}
  \end{minipage}
  \caption{Constraints on tensor-to-scalar ratio and running spectral
    index in correspondent extension of \emph{$\Lambda$CDM}
    model. Larger contours --- from \emph{WMAP7+BAO+$H_0$+SPT}
    dataset, smaller contours --- from \emph{CL+WMAP7+BAO+$H_0$+SPT}
    dataset.}
  \label{fig:atensnrun}
\end{figure}

It turns out, that existing cosmological data on galaxy cluster mass
function does not improve the constraints on tensor-to-scalar ratio
$r$ and running spectral index of primordial perturbations $dn_s/d
\ln(k)$ (see Fig.~\ref{fig:atensnrun}).

\section{Constraints on total neutrino mass and effective number of
  neutrino species}

\subsection{Massive neutrinos}
\label{sec:mnu}

Massive neutrinos with masses of order of few $0.1$\,eV would produce
significant suppression of the matter density fluctuation at galaxy
cluster scale since they remain to be relativistic after equipartition
and are started to be involved in gravitational instability growth
only later \citep[see, e.g.,][]{doroshkevich81,hu98}. This change the
relation between linear perturbations amplitude $\sigma_8$ and
normalization of CMB anisotropy power spectrum, which strongly depend
also on $\Omega_m$. If the measurement of Hubble constant is added to
the measurements of $\sigma_8$ and normalization of CMB anisotropy
power spectrum, the $\Omega_m$ and total neutrino mass are both
constrained with these data.

Since massive neutrinos are involved in gravitational instability
considerably later than dark matter particles and baryons, their
presence could change the process of non-linear gravitational collapse
of galaxy cluster haloes. The detailed analysis of this question using
numerical simulations show, that these changes are indeed notable, but
the correspondent change of cluster mass function appears to be not
larger than few percents for neutrinos with total mass about 0.3\,eV
and clusters with masses of order $10^{14}~M_\odot$
\citep{brandbyge10,marulli11}. This is smaller that the accuracy of
theoretical mass function calculations used in our work, which is
approximately equal to 5\% \citep{tinker08}.

\begin{figure}
  \centering
  \begin{minipage}{0.48\linewidth}
  \smfigure{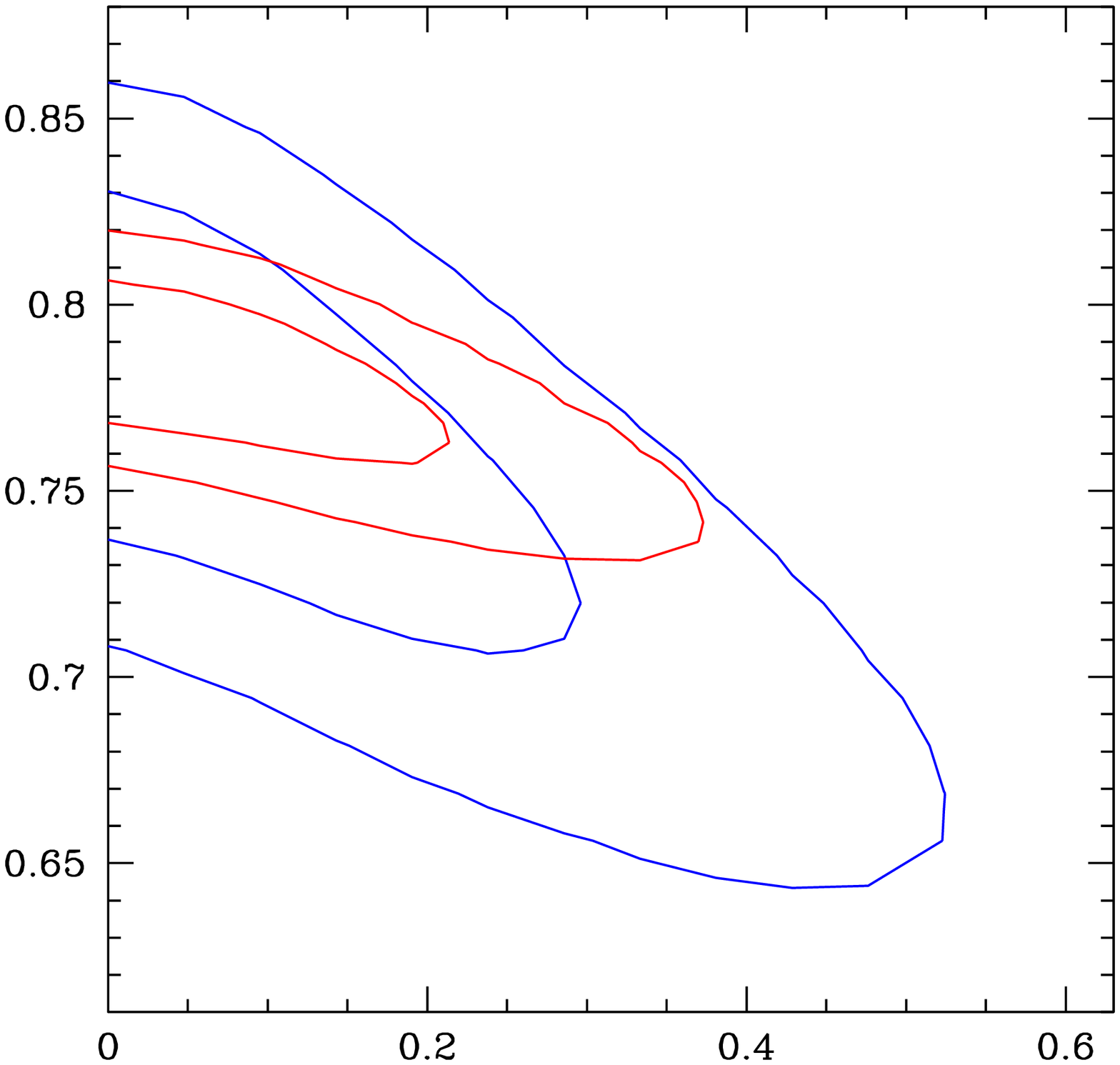}{$\Sigma m_\nu$, eV}{$\sigma_8$}
  \end{minipage}
  \begin{minipage}{0.48\linewidth}
    \smfigure{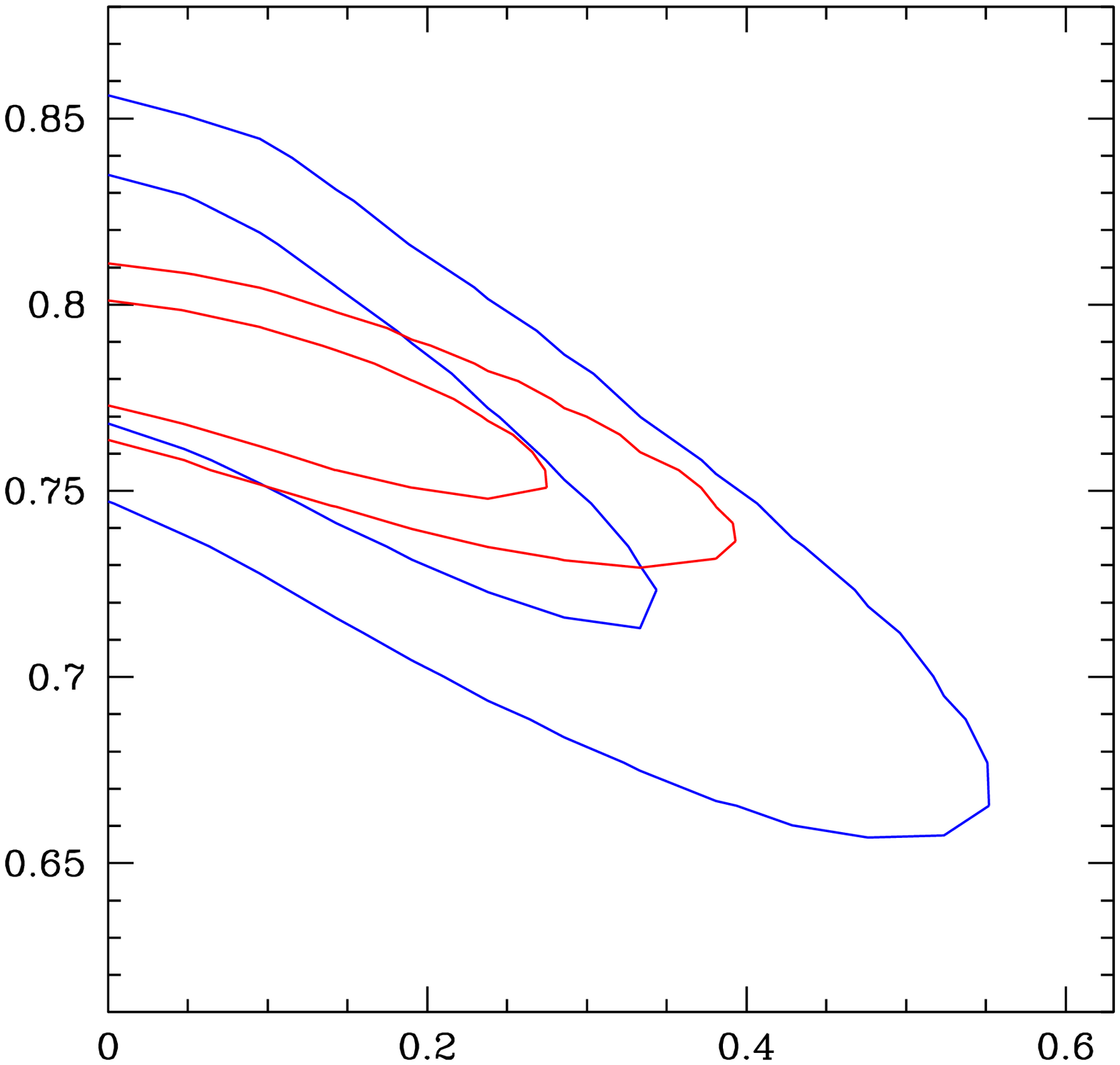}{$\Sigma m_\nu$, eV}{$\sigma_8$}
  \end{minipage}
  \caption{Constraints on total neutrino mass in \emph{$\Lambda$CDM}
    model with non-zero neutrino mass. Larger contours --- from
    \emph{WMAP7+$H_0$} dataset (left) and from
    \emph{WMAP7+BAO+$H_0$+SPT} dataset (right), smaller contours ---
    from the same data, with the data on galaxy cluster mass function
    added.}
   \label{fig:mnu}
\end{figure}

The constraints on total neutrino mass from \emph{CL+WMAP7+$H_0$}
dataset are shown in the left panel of Fig.~\ref{fig:mnu}, the upper
limit is $\Sigma m_\nu<0.29$~eV (95\% c. l., see also
Table~\ref{tab:pars}).  In the right panel of Fig.~\ref{fig:mnu} we
show the constrains, obtained with additional available data:
\emph{CL+WMAP7+BAO+$H_0$+SPT}, in this case the upper limit is $\Sigma
m_\nu<0.32$~eV. If the data on SN Ia are added, the constraint is
$\Sigma m_\nu<0.28$~eV. Therefore, the additional cosmological data
provide almost no changes in total neutrino mass constraints, as
compared to \emph{CL+WMAP7+$H_0$} dataset.

From these results we see, that new cosmological data does not
significantly improve the constraints on total neutrino mass, which
were obtained from the same galaxy cluster data in \cite{av09b} ---
$\Sigma m_\nu<0.33$~eV. Among the other recent constraints on total
neutrino mass, we note the upper limit $\Sigma m_\nu<0.58$\,eV,
obtained using the data on the CMB anisotropy spectrum and Hubble
constant measurements \citep{komatsu11}. Also significant constraints
were obtained from the other data on cluster mass function or density
perturbations power spectrum
\citep{malinovsky08,mantz10b,reid10,thomas10}. The strongest upper
limits on total neutrino mass was published in papers, based on the
observations of Lyman-$\alpha$ forest \citep{seljak06}. However, the
more accurate treatment of systematic errors should weaken these
constraints considerably \citep{bolton08}. A recent review of total
neutrino mass measurements from astrophysical data can be found in
\cite{abazajian11a}.

We note, that all the constraints on total neutrino mass described
above were obtained in assumption that only photons and three known
neutrino species are relativistic at matter--radiation equipartition
epoch. It turns out that if additional neutrino species are allowed in
the model, the constraints on total neutrino mass change significantly
(see below).

\subsection{Effective number of neutrino species}
\label{sec:neff}

The present CMB radiation density is directly related to the density
of relativistic matter at matter--radiation equipartition epoch, only
if the number of relativistic particles species at equipartition is
assumed to be known --- a set of relativistic particles consists of
photons and three known neutrino species. If the number of
relativistic particles at equipartition differ from its standard
value, than the relativistic matter density is not defined anymore. In
this case the determination of relativistic matter density at
equipartition from cosmological data gives the measurement of the
effective number of relativistic species.

The number of relativistic species is usually parametrized with the
number of species of light neutrinos $\neff$, which gives the
following relation between the densities of relativistic mater and CMB
photon energy: 
$$\rho_r=\left[1+\frac{7}{8}\left(\frac{4}{11}\right)^{4/3}\neff\right]\rho_\gamma$$
Note, that for the case of three known neutrino species accurate
calculations of neutrino decoupling give somewhat larger effective
number of neutrino species, $\neff=3.046$
\citep[e.g.,][]{gnedingnedin98,dolgov99,mangano02}. 

One of the main observables, which is measured from the CMB
observations is the size of particle horizon at radiation--matter
energy density equipartition and, therefore, equipartition redshift
$z_{eq}$ \citep[e.g.,][]{hudod02,gorbunov10}. On the other hand, as it
was shown above, galaxy cluster mass function data allow to
significantly improve the measurement of $\Omega_m h^2$, using the
measurement of the matter density perturbation amplitude $\sigma_8$
and CMB anisotropy normalization. The measurement of two parameters,
$z_{eq}$ and $\Omega_m h^2$, gives the measurement of relativistic
energy density at equipartition, which allow to obtain constraints on
$\neff$.

\begin{figure}
  \centering
  %    \smfigure{plot_2d_nrel_omegamh2.ps}{$\neff$}{$\Omega_m h^2$}
  \begin{minipage}{0.48\linewidth}
    \smfigure{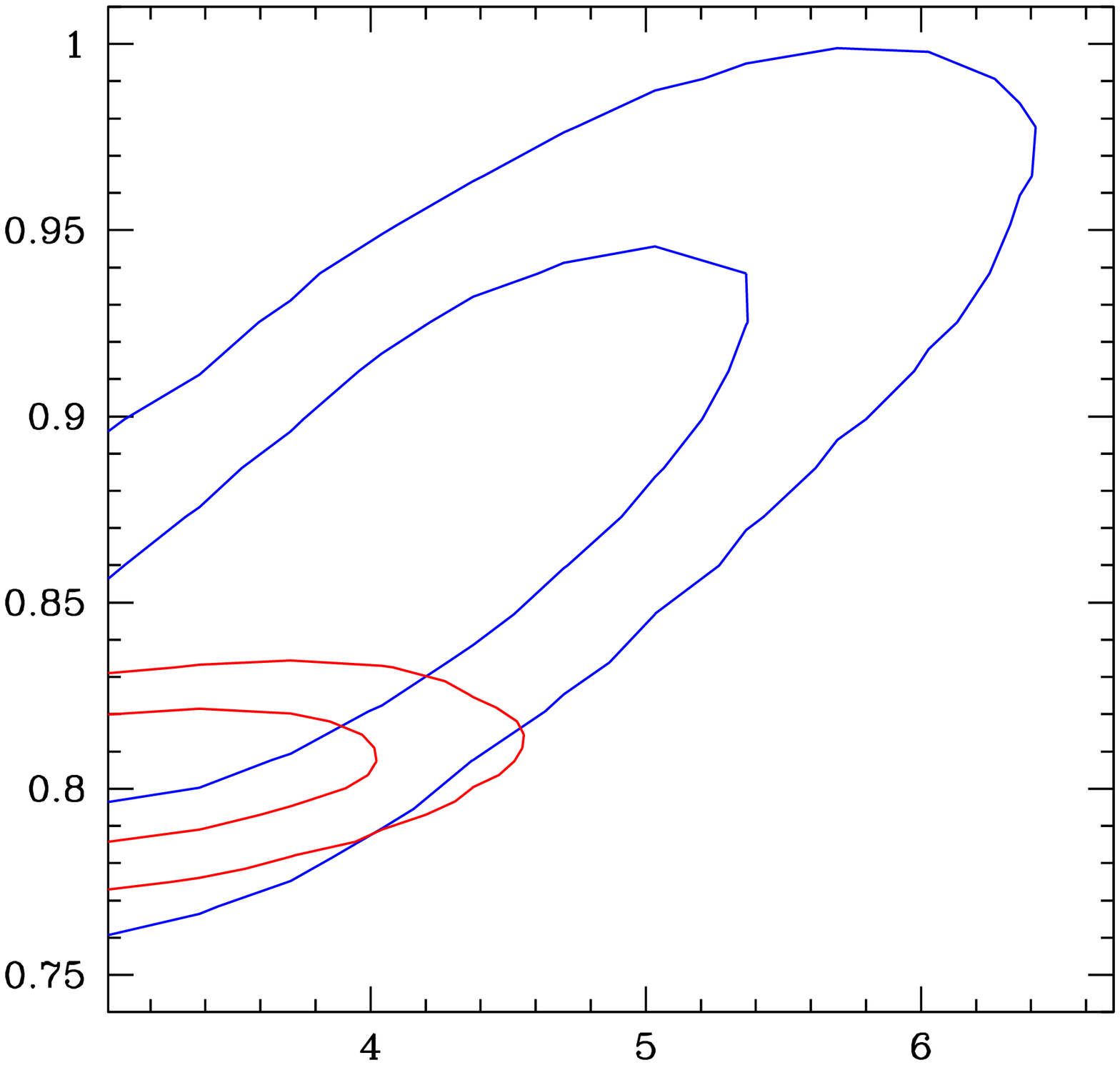}{$\neff$}{$\sigma_8$}
  \end{minipage}
  \begin{minipage}{0.48\linewidth}
    \smfigure{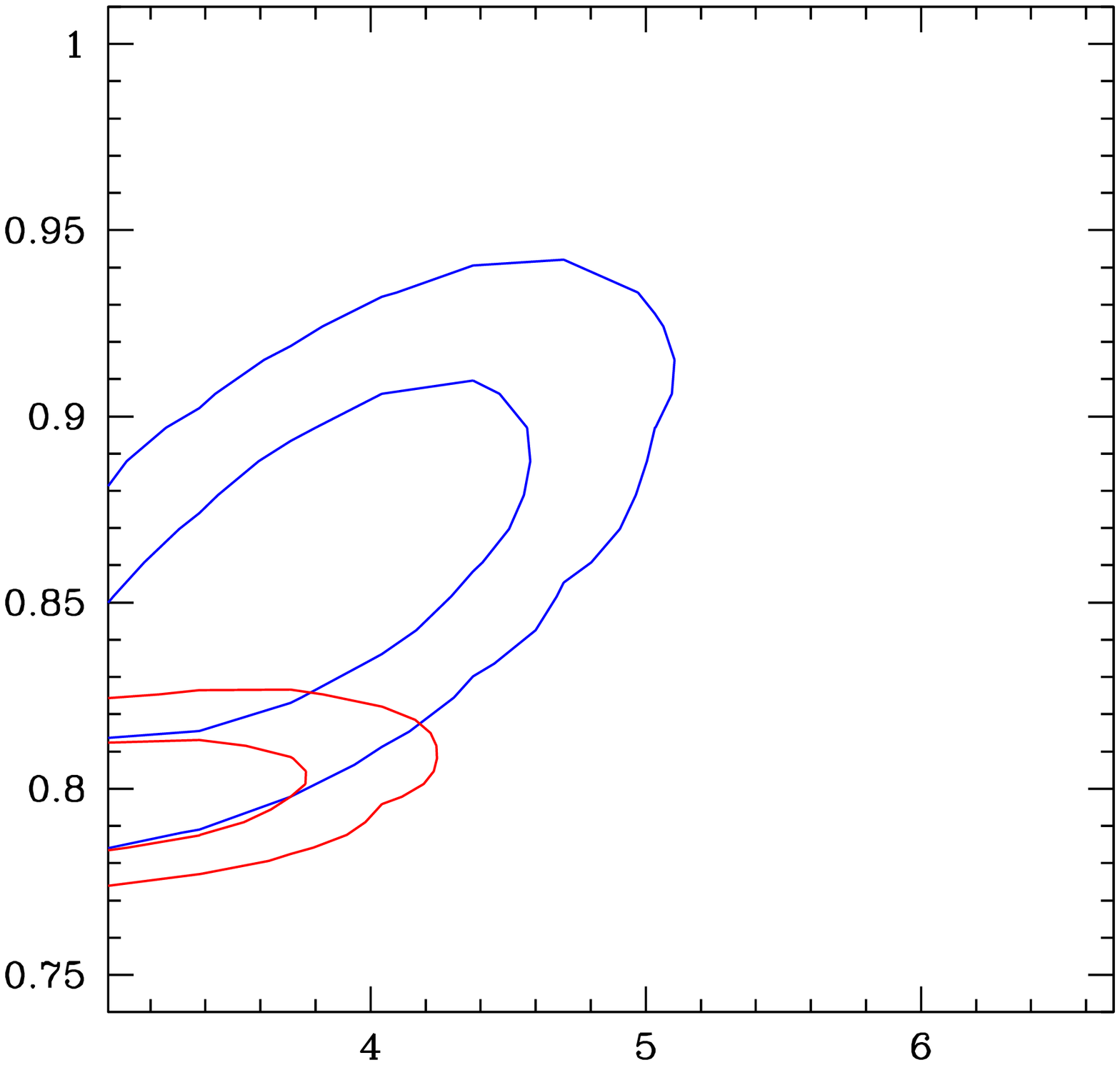}{$\neff$}{$\sigma_8$}
  \end{minipage}
  
  \caption{Constraints on the effective number of neutrino species in
    \emph{$\Lambda$CDM} model with free $\neff$. Larger contours ---
    from \emph{WMAP7+BAO+$H_0$} dataset (left) and from
    emph{WMAP7+BAO+$H_0$+SPT} (right), smaller contours --- from the
    same data, with the data on galaxy cluster mass function added.}
  \label{fig:nrel}
\end{figure}

The constraints on the effective number of light neutrino species
obtained from joint analysis of data on galaxy cluster mass function
and \emph{WMAP7+BAO+$H_0$} dataset is shown in left panel of
Fig.~\ref{fig:nrel}, the upper limit is $\neff<4.07$.  If the data on
CMB power spectrum at higher multipoles are added, the constraint is
improved, because the damping of CMB power at small angular scales
depends on $\neff$ as well. This is shown in the right panel of
Fig.~\ref{fig:nrel}, where \emph{SPT} data are added to
\emph{WMAP7+BAO+$H_0$} dataset. With these data taken in account, the
upper limit is $\neff<3.74$. Note, that the data on galaxy cluster
mass function not only improve the constraints on $\neff$, but it is
also shift them closer to the standard value $\neff=3.046$. The data
on SN Ia also improve this constraints slightly, in this case we
obtain $\neff<3.70$. Systematical errors in cluster mass measurements
make these upper limits slightly less strong. For example, if cluster
masses are underestimated for $\delta M/M = 0.09$, from
\emph{CL+WMAP7+BAO+$H_0$+SPT} dataset we obtain $\neff< 3.89$.

These results can be compared to the following constraints obtained
recently. Using the results of Atacama Cosmology Telescope
(\emph{ACT}) observations in combination with the data on $H_0$ and
BAO measurements, the following constraint was obtained:
$\neff=4.56\pm0.75$ \citep[68\% c.l.,][]{dunkley10}. Using new data of
South Pole Telescope (\emph{SPT}) survey together with the same
measurements of $H_0$ and BAO, the constraint $\neff=3.86\pm0.42$ was
obtained \citep{keisler11}. In these constraints somewhat higher
values of $\neff$ are preferred, but standard value $\neff=3.046$ is
not ruled out at 95\% confidence level.

In \cite{keisler11}, \emph{SPT} data were analyzed jointly with the
data on galaxy cluster mass function in the form of constraint on
$\sigma_8(\Omega_M /0.25)^{0.47}$ parameter combination, taken from
\cite{av09b}. In our work stronger constraints on $\neff$ were
obtained since we used complete likelihoods for galaxy cluster mass
function data. Note also that, as compared to \cite{keisler11},
systematic errors are not included in our upper limits on $\neff$ (see
discussion above).

Compatible constraints on $\neff$ were obtained from the other data on
galaxy cluster mass function. For example, in \citep{mantz10b} the
constraint $\neff=3.4^{+0.6}_{-0.5}$ (68\% c.l.) was
obtained. Therefore, the constraints from the data on galaxy clusters
are in better agreement with the standard value $\neff=3.046$, as
compared to \emph{ACT} and \emph{SPT} results. A significant
constraint, which is also consistent with the standard value
$\neff=3.046$, was obtained from the comparison of observed abundance
of light elements with the predictions of primordial nucleosynthesis
theory \citep{mangano11}.

We emphasize that our constraints on $\neff$, and also all other
constraints on this parameter, discussed above (except the constraint
from observed light element abundance), were obtained in assumption of
zero total neutrino mass. However, if both total neutrino mass and
$\neff$ are considered as free parameters, the constraints on both of
them turn to be substantially weaker. This case is considered below.

\begin{figure}
  \centering
  \begin{minipage}{0.48\linewidth}
    \smfigure{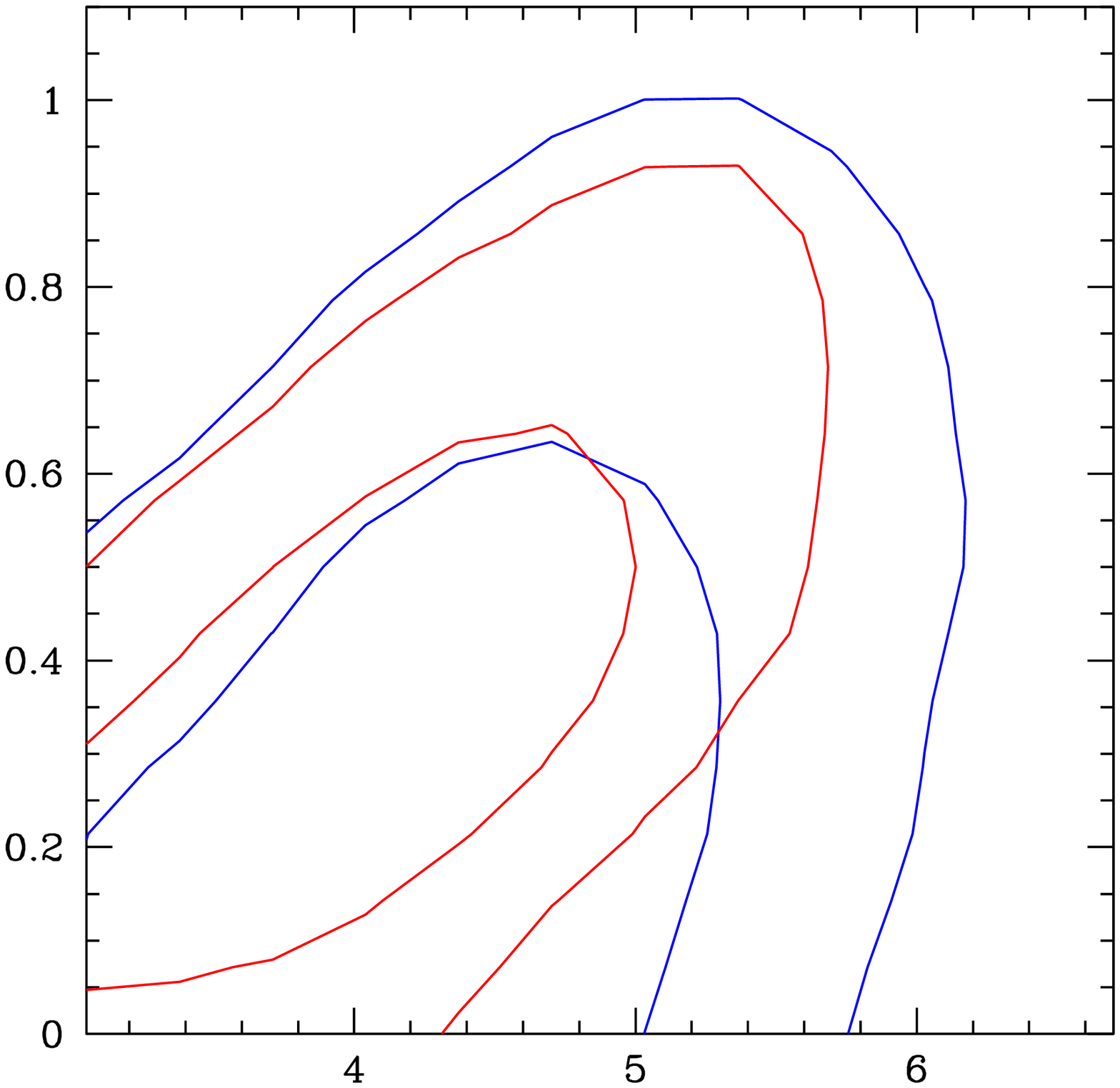}{$\neff$}{$\Sigma m_\nu$, eV}
  \end{minipage}
  \begin{minipage}{0.48\linewidth}
    \smfigure{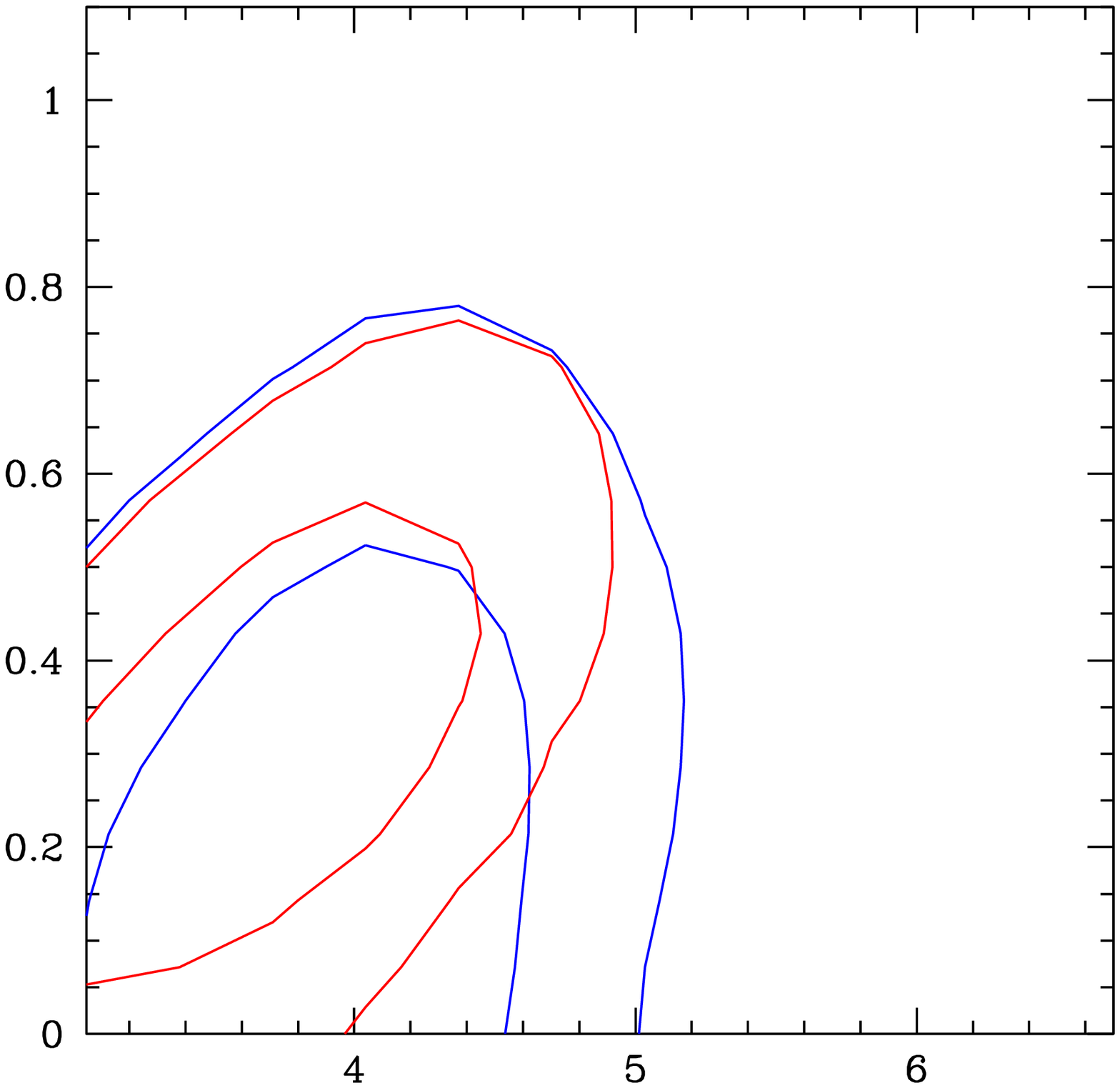}{$\neff$}{$\Sigma m_\nu$, eV}
  \end{minipage}
  
  \caption{Constraints on effective number of neutrino species and on
    total neutrino mass in \emph{$\Lambda$CDM} model with free $\neff$
    and $\Sigma m_\nu$. Larger contours correspond to
    \emph{WMAP7+BAO+$H_0$} dataset (left) and
    \emph{WMAP7+BAO+$H_0$+SPT} dataset (right), smaller contours ---
    to the same data, with the data on galaxy cluster mass function
    added.}
  \label{fig:nnumnu}
\end{figure}

\subsection{Effective number of massive neutrinos}
\label{sec:mnunrel}

\begin{figure}
  \centering
  \begin{minipage}{0.48\linewidth}
    \smfigure{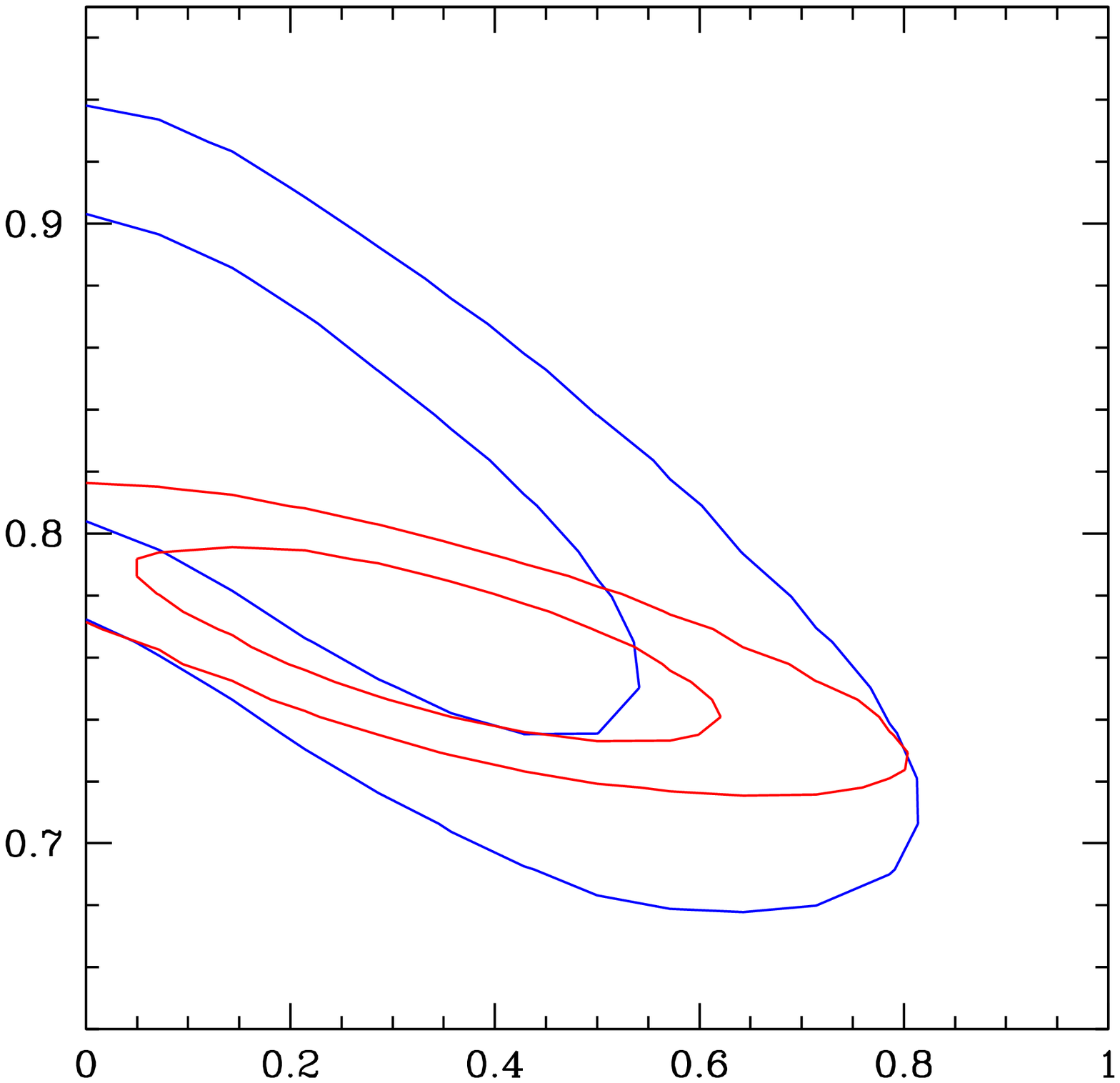}{$\Sigma m_\nu$, eV}{$\sigma_8$}
  \end{minipage}
  \begin{minipage}{0.48\linewidth}
    \smfigure{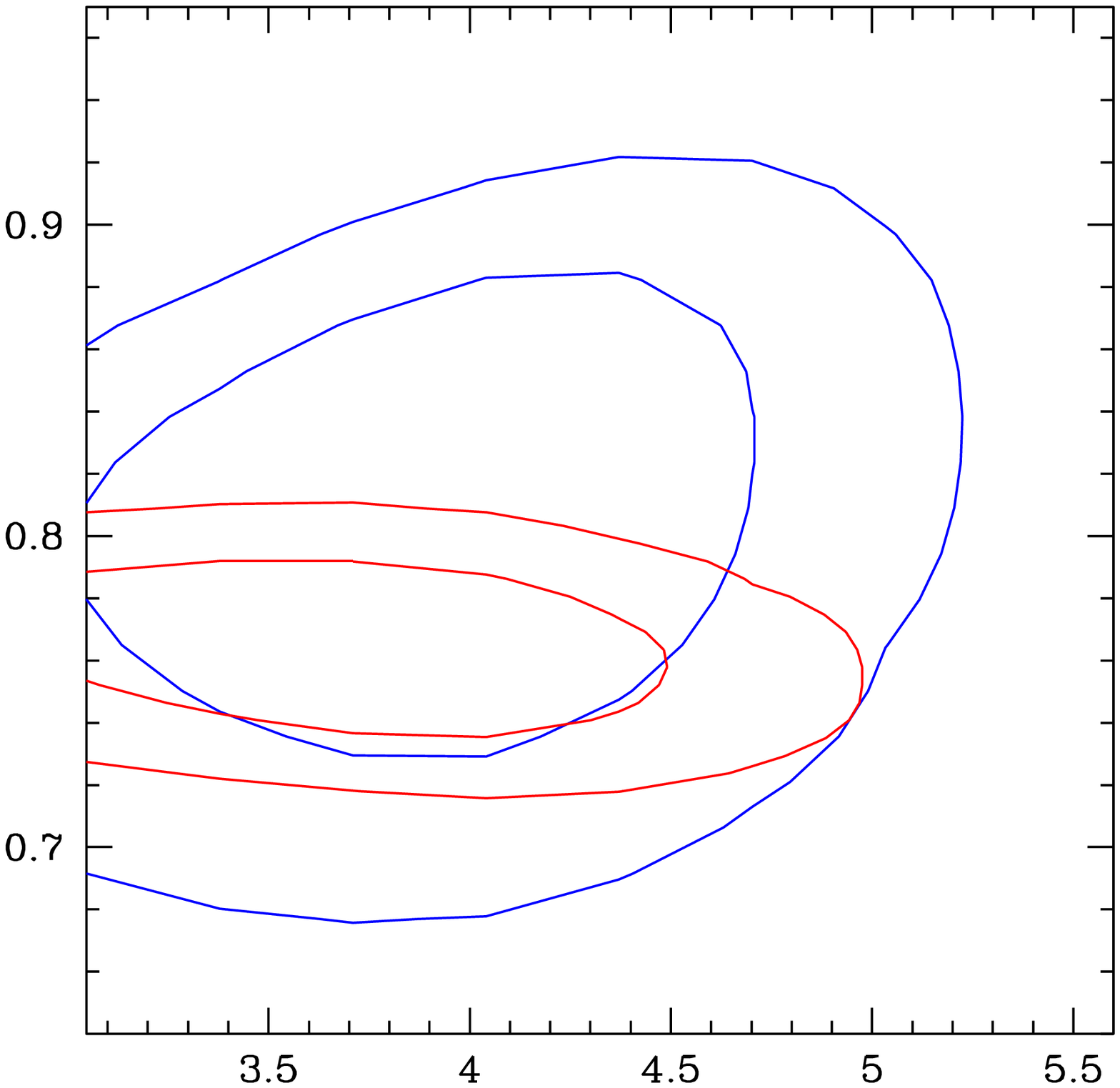}{$\neff$}{$\sigma_8$}
  \end{minipage}
  \caption{Constraints on effective number of neutrino species and on
    total neutrino mass in \emph{$\Lambda$CDM} model with free $\neff$
    and $\Sigma m_\nu$. Larger contours correspond to
    \emph{WMAP7+BAO+$H_0$+SPT} dataset, smaller contours --- to the
    same data, with the data on galaxy cluster mass function added.}
  \label{fig:nnumnu2}
\end{figure}

The constraints on total neutrino mass and effective number of
neutrino species $\neff$ in \emph{$\Lambda$CDM} model with arbitrary
number of massive neutrinos are shown in Fig.~\ref{fig:nnumnu} and
\ref{fig:nnumnu2} (see also Table~\ref{tab:pars}). Note, that if
arbitrary number of neutrino species are allowed in the model, the
upper limits on their total mass turn to be substantially weaker. And
vice-versa, the assumption on non-zero total neutrino mass weaken the
upper limits on number of neutrino species. We obtain the following
upper limits from \emph{CL+WMAP7+BAO+$H_0$+SPT} dataset: $\Sigma
m_\nu<0.72$\,eV and $\neff<4.62$.

Therefore, the most powerful set of cosmological data used in our work
(\emph{CL+WMAP7+BAO+$H_0$+SPT}) does not exclude the existence of one
additional type of neutrino. Moreover, for these data the maximum
likelihood is shifted to the model with total neutrino mass about
$0.4$\,eV and number of neutrino species $\neff\approx4$. The
improvement of maximum likelihood for this model, as compared to the
model with $\Sigma m_\nu = 0$ and $\neff=3.046$, is $\Delta \ln L =
1.80$, which corresponds to $\Delta \chi^2 = 3.60$ for 2 degrees of
freedom, approximately $1.4\sigma$ significance. From
Fig.~\ref{fig:nnumnu2} one can see that the likelihood is improved
mainly due to the assumption on non-zero neutrino mass. When massive
neutrinos are added to the model with free number of neutrino species,
the likelihood improvement corresponds to $\Delta \chi^2 = 3.41$ for
one degree of freedom, which corresponds to approximately $1.9\sigma$
significance.

Systematical errors of cluster mass measurements decrease the
significance of introduction of these new parameters in the model. For
example, if cluster masses are underestimated for $\delta M/M = 0.09$,
then the introduction of massive neutrinos into the
\emph{$\Lambda$CDM} model with free $\neff$ gives $\Delta \chi^2 =
2.83$ and significance about $1.7\sigma$. One can see, that systematic
uncertainties in cluster mass measurements are very significant in
total neutrino mass constraints even with existing cluster
data. Therefore, in order to improve similar measurements in future
one will need to significantly reduce systematic uncertainties in
cluster mass measurements.

\subsection{Different mass distributions of neutrino species}

All constraints on total neutrino mass, discussed above, were obtained
in assumption that three known neutrino species have the same equal
masses (and additional neutrinos are massless), i.e.\ the mass
distribution of these three known neutrino species is
degenerate. However, from the observations of neutrino oscillation it
is known, that masses of different neutrino species are also
different. Generally, if the mass distribution of neutrino species is
non-degenerate, the constraints on total neutrino mass from
cosmological data may change. This happens for several reasons
\citep[see, e.g.,][]{slosar06}, in particular since most massive
neutrino species become non-relativistic earlier and spectrum of
linear density perturbations is suppressed at smaller scales. Also, if
neutrino mass is larger then approximately 0.5\,eV, they become
non-relativistic before recombination and significantly change angular
power spectrum of CMB anisotropy.

In case of $\neff=3.046$, the assumption on non-degenerate neutrino
mass distribution makes almost no change in the neutrino mass
constraints from cosmological data. For example, if there is only one
massive neutrino, the constraint on total neutrino mass from
\emph{CL+WMAP7+BAO+$H_0$+SPT} dataset weaken only very slightly,
$\Sigma m_\nu<0.33$\,eV (as compared to $\Sigma m_\nu<0.32$\,eV, in case
of three degenerate neutrinos, see above). This is quite consistent
with what is expected for similar cosmological data
\citep[e.g.,][]{takada06,slosar06}.

\begin{figure}
  \centering
  \begin{minipage}{0.48\linewidth}
    \smfigure{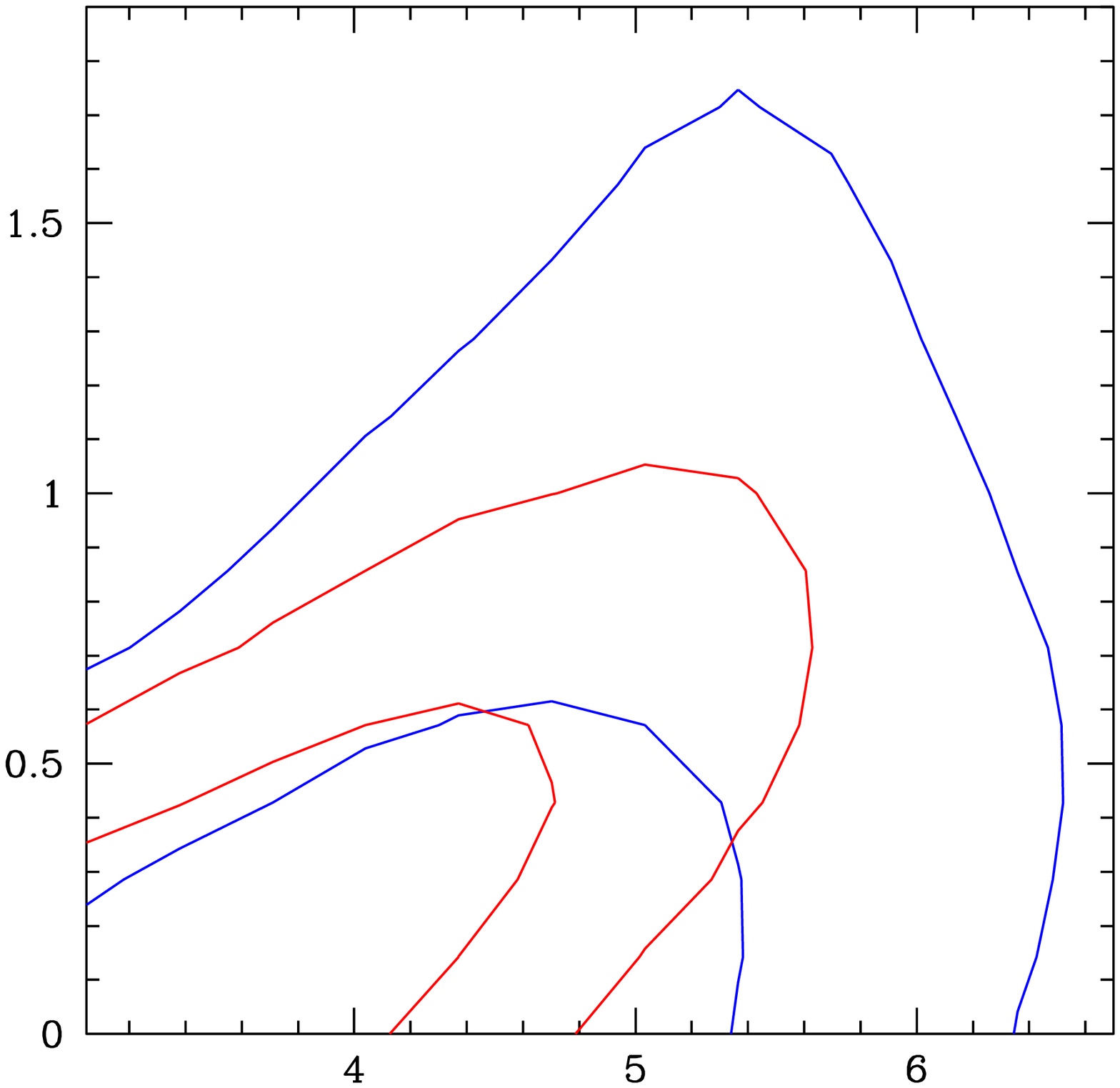}{$\neff$}{$\Sigma m_\nu$, eV}
  \end{minipage}
  \begin{minipage}{0.48\linewidth}
    \smfigure{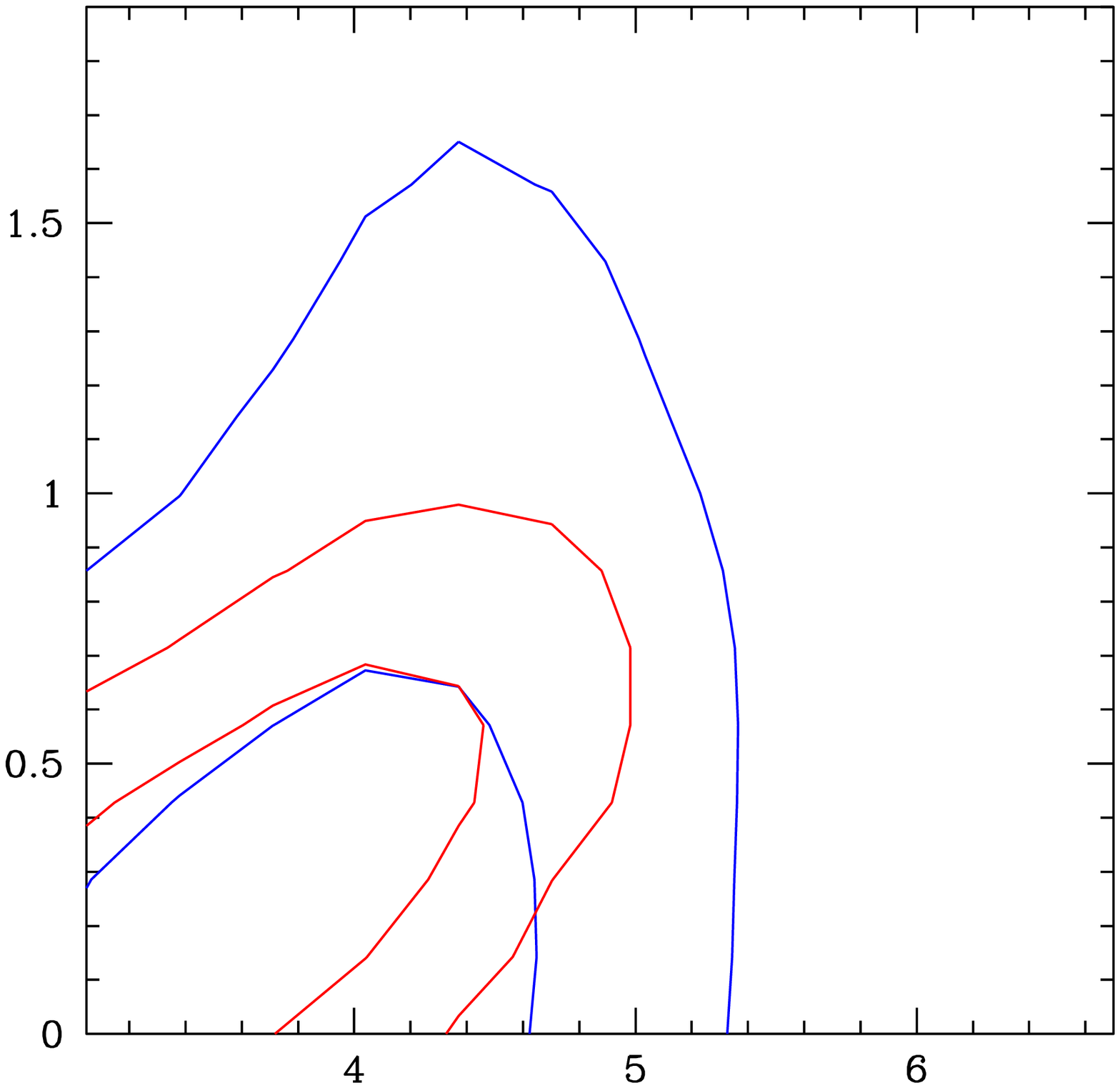}{$\neff$}{$\Sigma m_\nu$, eV}
  \end{minipage}
  
  \caption{The same constraints as in Fig.~\ref{fig:nnumnu}, but in
    assumption that there is only one type of neutrino with non-zero
    mass.}
  \label{fig:nnumnum1}
\end{figure}

It turns out, that with increasing number of neutrino species, the
assumption that only one neutrino is massive weaken the constraint on
total neutrino mass more significantly. This is shown in
Fig.~\ref{fig:nnumnum1} --- the upper limit for
\emph{CL+WMAP7+BAO+$H_0$+SPT} dataset is $\Sigma m_\nu<0.88$\,eV in
this case. The upper limit on number of neutrino species changes only
slightly and is $\neff<4.68$. These upper limits can be compared to
$\Sigma m_\nu<0.72$~eV and $\neff<4.62$ in case of three degenerate
neutrinos (see above).

\subsection{Light sterile neutrinos}

The constraints on $\Sigma m_\nu$ and $\neff$ discussed above can have
important application to study the possibility of existence of light
sterile neutrinos with masses near $1$\,eV, which were suggested in
order to explain the results of short baseline neutrino oscillations
searches experiments \emph{LSDN} \citep{aguilar01} and
\emph{MiniBooNE} \citep{aguilararevalo10}, and also recently
discovered, so-called reactor neutrino anomaly
\citep{mueller11,mention11}. In order to explain these results a
possibility of the existence of one or two light sterile neutrino
species \citep{maltoni07,karagiorgi09,melchiorri09,akhmedov10,kopp11},
which should be thermalized in early Universe \citep[see, e.g.,
discussion and references in][]{hamann10} was discussed. The presence
of these additional neutrinos should also change the value of $\neff$,
measured from cosmological data.

The upper limit $\neff<3.70$, which was obtained in our work in
assumption on zero total neutrino mass (see above), does not agree
with the existence of even one additional type of neutrino. However,
in order to explain the results of short baseline neutrino
oscillations experiments additional species of \emph{massive}
neutrinos are suggested \citep[e.g.,][]{mention11,kopp11}. In this
case the constraints on $\neff$ turn to be significantly weaker. As it
was shown above, all considered cosmological data are consistent with
the existence of one additional type of light sterile neutrino with
mass about $0.4$\,eV.

It was discussed recently, that the results of short baseline neutrino
oscillations searches are better explained if two additional light
sterile neutrino species are suggested
\citep{kopp11,giunti11a}. However, recent improvements of the
\emph{MiniBooNE} experiment data, probably will allow to explain all
the data of these experiments with only one additional light sterile
neutrino \citep{giunti11b}. Anyway, in order to explain the results of
these experiment assuming only one additional type of neutrino, the
masses near $1$\,eV are probably required
\citep{kopp11,giunti11b}. This value disagree with upper limits on
total neutrino mass, obtained above --- even in most conservative case
of one additional massive neutrino the upper limit on total neutrino
mass is $\Sigma m_\nu<0.88$\,eV at 95\% confidence level. Therefore, we
conclude that masses of light sterile neutrinos, which may explain
current results of short baselines neutrino oscillation searches are
not in good agreement with our constraints.

In order to reconcile neutrino masses near $1$\,eV with the
constraints on total neutrino mass from cosmological data it was
suggested to consider the cosmological models with dark energy more
general than cosmological constant \citep{kristiansen11,hamann11}. It
was found that with dark energy equation of state parameter $w<-1$ the
limits on neutrino mass are relaxed considerably. This remains true
also when the galaxy cluster mass function cosmological data taken in
account. The upper limits on total neutrino mass and effective number
of neutrino species in the model with free $w$, and free number of
massive neutrinos are $\Sigma m_\nu<0.85$~eV and $\neff<4.33$ in
assumption that three neutrino species have equal non-zero mass, and
$\Sigma m_\nu<1.00$~eV and $\neff<4.39$ in case if there is only one
type of massive neutrino in the model. Therefore, the limits on total
neutrino mass, obtained with galaxy cluster mass function data taken
in account, are still in poor agreement with suggested sterile
neutrino mass near $1$\,eV, even in cosmological models with dark
energy equation of state parameter $w<-1$.

Therefore, all available cosmological data are consistent with one
additional light sterile neutrino. However, their total mass may be
near $0.4$\,eV, while the values of total neutrino mass near $1$\,eV
are in poor agreement with existent cosmological data. We emphasize
that the constraints obtained in our work apply only to light sterile
neutrinos thermalized in early Universe.
\begin{figure}
  \centering
  \begin{minipage}{0.49\linewidth}
    \smfigure{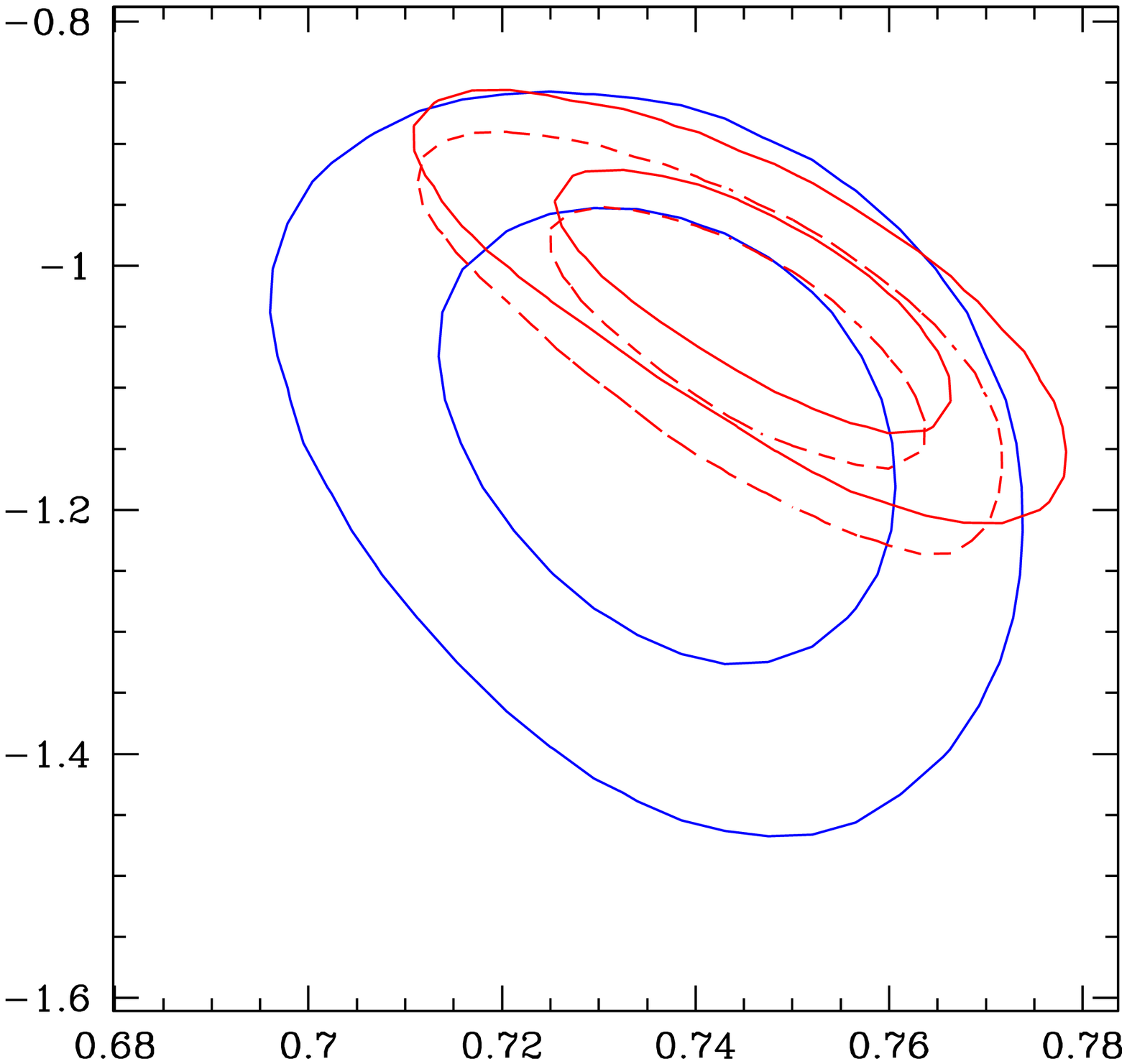}{$\Omega_\Lambda$}{$w$}
  \end{minipage}
  \begin{minipage}{0.49\linewidth}
    \smfigure{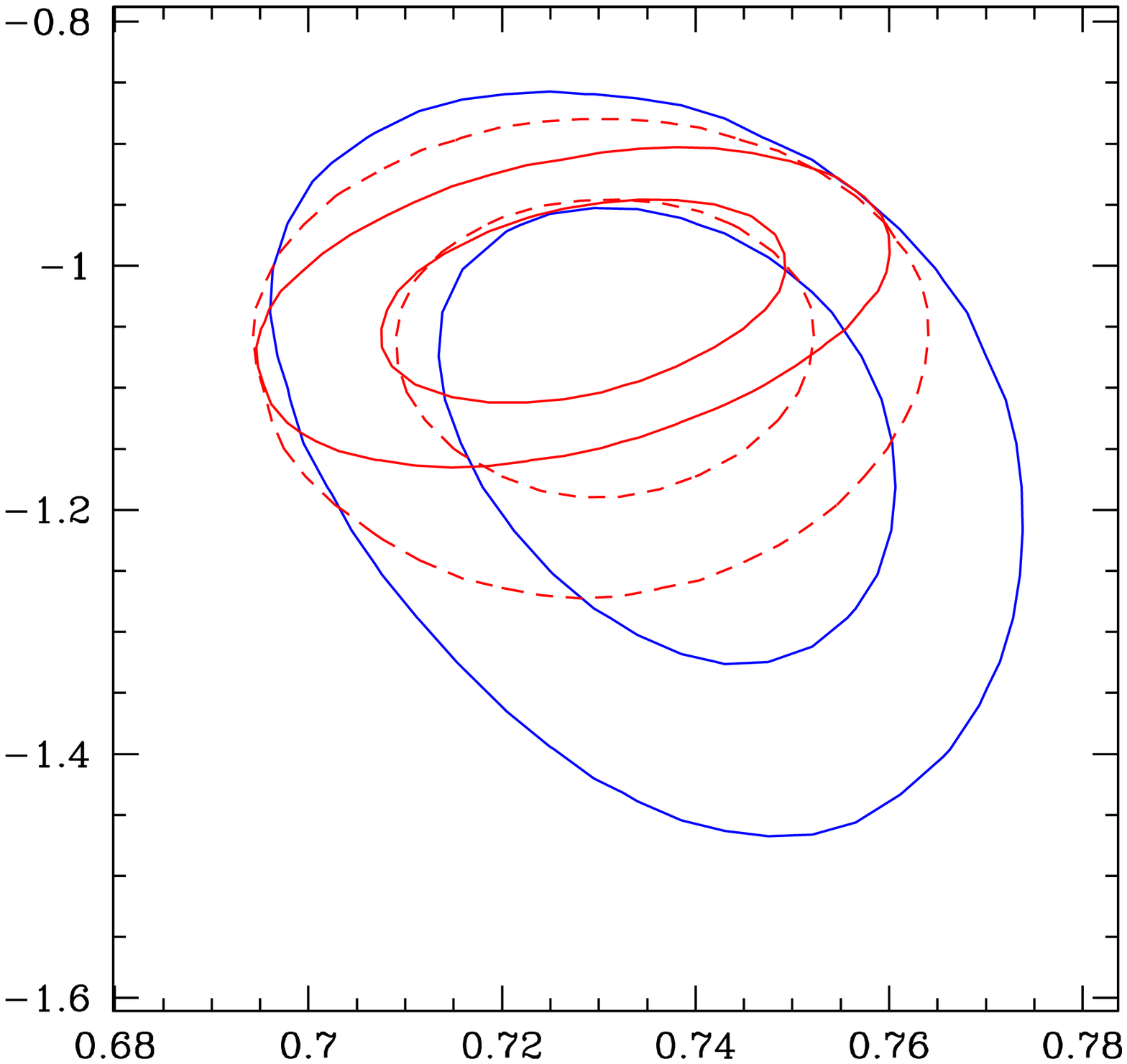}{$\Omega_\Lambda$}{$w$}
  \end{minipage}
  \caption{The constraints on dark energy density and equation of
    state parameter in \emph{WCDM} model. Larger contours --- for
    \emph{WMAP7+BAO+$H_0$} dataset, smaller contours --- for
    \emph{WMAP7+BAO+$H_0$+CL} dataset (left panel) and for
    \emph{WMAP7+BAO+$H_0$+SN} dataset (right panel). Dashed line in
    left panel show the contours for systematic $\delta M/M\approx
    0.09$ shift in cluster mass measurements. In the right panel the
    contours for SN Ia data with systematic uncertainties taken in
    account are shown with dashed line.}
  \label{fig:olwsncl}
\end{figure}

\section{Dark energy constraints}
\label{sec:de}

In order to obtain most powerful constraints on dark energy equation
of state, the measurements of accelerated expansion of Universe from
the observations of distant supernovae type Ia were used in
\cite{komatsu11}. On the other hand, during the last years it become
clear that the errors of cosmological measurements based on SN Ia data
are dominated by systematic uncertainties \citep{hicken09,kessler09}.
In recent works on SN Ia observations, systematical errors are
estimated to be very significant and to be larger than statistical
errors in these data. These uncertainties are mostly consist of
uncertainties of photometrical calibrations, color corrections on the
extinction in host galaxies, selection effects in SN Ia surveys and
others \citep[e.g.,][]{amanullah10}.

With these considerations in mind, it is interesting to study the
possibilities to obtain powerful dark energy constraints independently
on SN Ia measurements. We show below that these constraints may be
obtained using galaxy cluster mass function measurements, and their
uncertainties are not currently dominated by systematics. The updated
dark energy constraints obtained from the combination of recent
cosmological data are also given below.

\subsection{Flat Universe}
\label{sec:flatu}

\begin{figure}
  \centering
  \begin{minipage}{0.9\linewidth}
    \smfigure{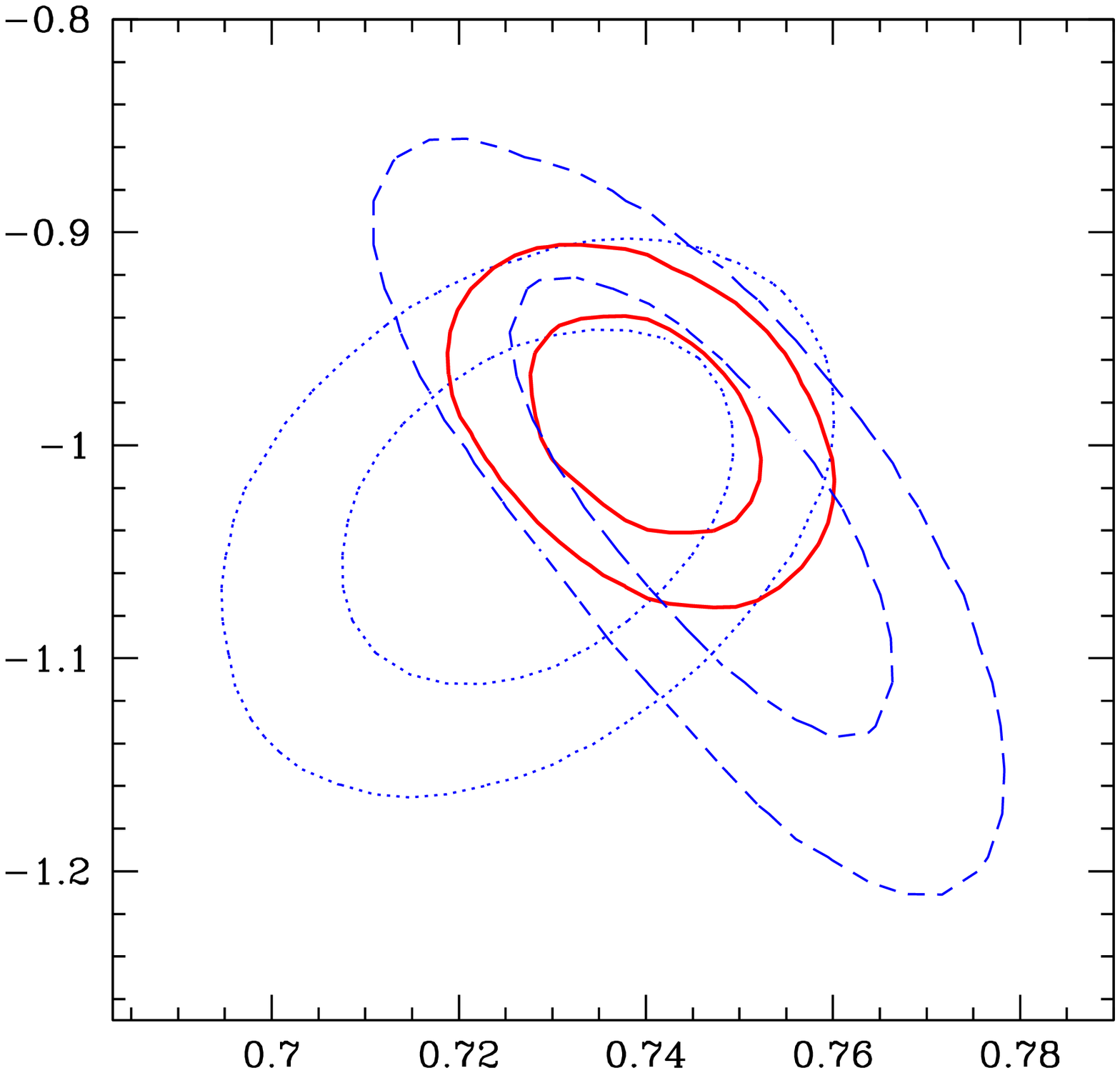}{$\Omega_\Lambda$}{$w$}
  \end{minipage}
  \caption{The constraints on dark energy density and equation of
    state parameter in \emph{WCDM} model. Larger contours --- for
    \emph{WMAP7+BAO+$H_0$+SN} dataset (dotted lines) and also for
    \emph{WMAP7+BAO+$H_0$+CL} (dashed lines), smaller contours --- for
    all the data (\emph{WMAP7+BAO+$H_0$+SN+CL}) combined.}
  \label{fig:olwsncl2}
\end{figure}

The constraints on the density and equation of state parameter $w$ of
dark energy in a flat Universe with free $w$ (\emph{WCDM} model) are
shown in Fig.~\ref{fig:olwsncl} and \ref{fig:olwsncl2} (see also
Table~\ref{tab:pars}). With no systematic uncertainties taken in
account, the existing data on galaxy cluster mass function give
somewhat weaker constraints, as compared to the data on SN Ia.
However, the error on $w$ parameter measured from from SN Ia data is
dominated by systematic uncertainties (see right panel of
Fig.~\ref{fig:olwsncl}). With systematical errors taken in account,
the constraint on dark energy equation of state parameter from
\emph{WMAP7+BAO+$H_0$+SN} dataset is $w = -1.068\pm 0.077$. On the
other hand, the \emph{WMAP7+BAO+$H_0$+CL} dataset gives the following
constraint: $w = -1.026\pm 0.069$~(stat.)~$\pm\,0.028$~(sys.), i.e.,
the total error is $\pm0.074$ in this case. Therefore, the constraints
from galaxy cluster mass function data and from SN Ia observations are
comparably powerful. Note that, in contrast to SN Ia data, the errors
from cluster mass function measurements are not currently dominated by
systematic uncertainties.

The constraints for all the data combined are shown in
Fig.~\ref{fig:olwsncl2}. One can see that galaxy cluster mass function
data significantly improve the dark energy equation of state parameter
constraints. The reason is that these data are independent and have
different degeneracies in parameter space.  From these data we obtain
the following measurement: $w = -0.990\pm
0.034$~(stat.)~$\pm\,0.041$~(sys.), where SN Ia systematic
uncertainties are also included in resulting systematic error. These
constraints appears to be somewhat better than those obtained in
\cite{av09b}.

\begin{figure}
  \centering
  \begin{minipage}{0.49\linewidth}
    \smfigure{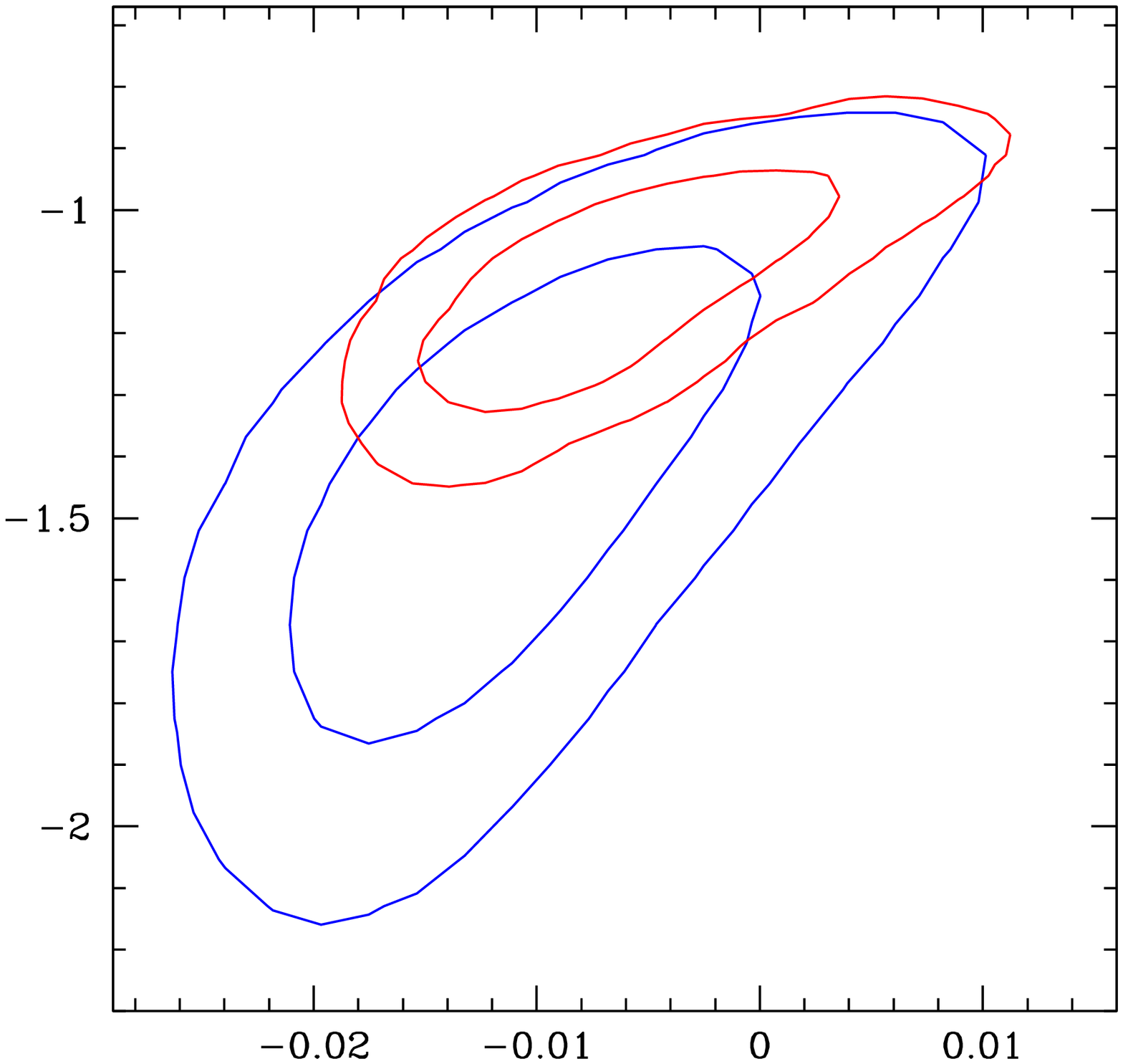}{$\Omega_k$}{$w$}
  \end{minipage}
  \begin{minipage}{0.49\linewidth}
    \smfigure{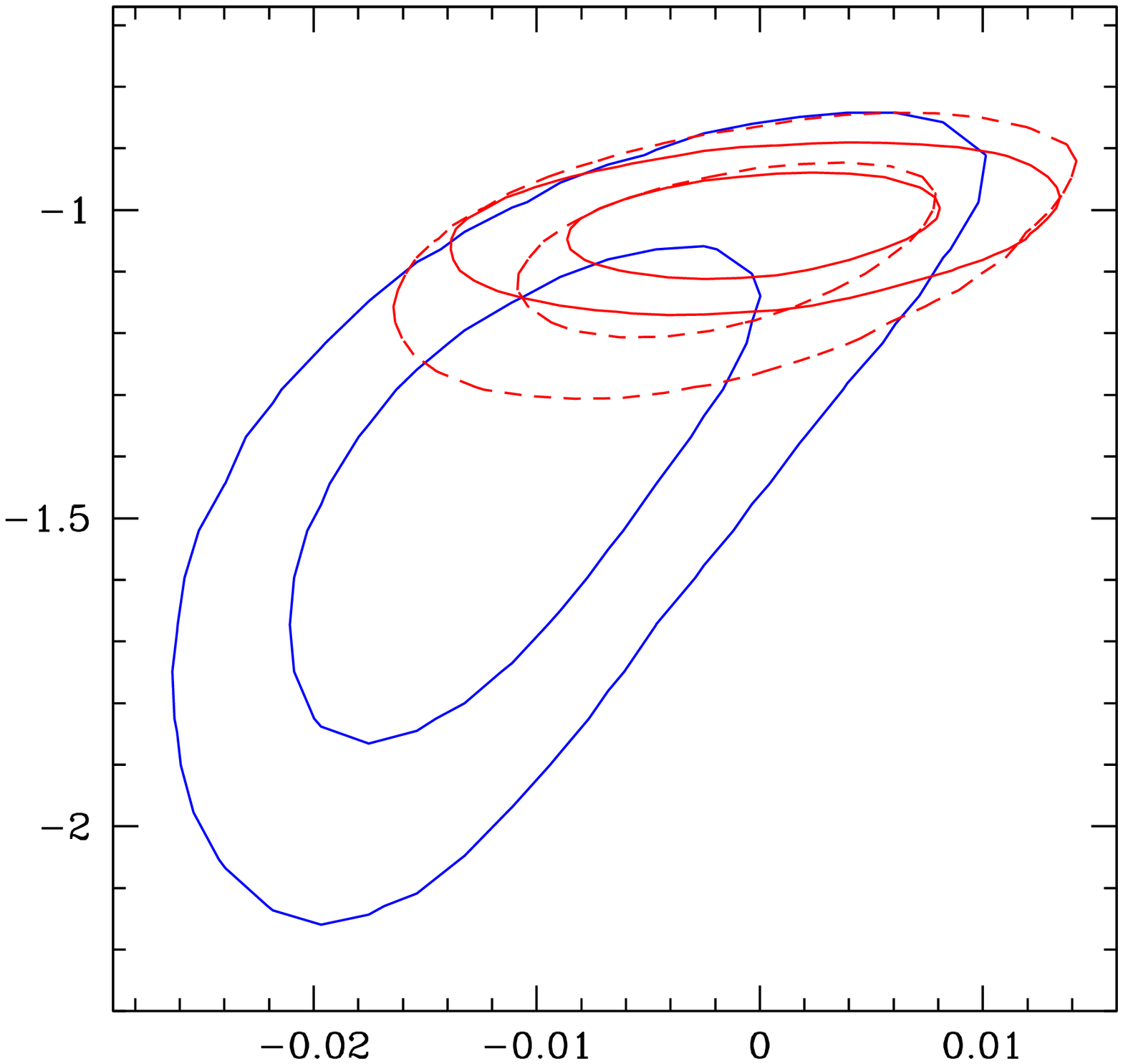}{$\Omega_k$}{$w$}
  \end{minipage}
  \caption{The constraints on the curvature of Universe and dark
    energy equation of state parameter in \emph{WCDM} model with
    $\Omega_k\ne 0$. Larger contours --- for \emph{WMAP7+BAO+$H_0$}
    dataset, smaller contours --- for the same data with the data on
    clusters (left panel) and SN Ia (right panel) added. The contours
    for SN Ia data with systematic uncertainties taken in account are
    shown with dashed lines in right panel of the Figure.}
  \label{fig:okwsncl}
\end{figure}
\begin{figure}
  \centering
  \begin{minipage}{0.9\linewidth}
    \smfigure{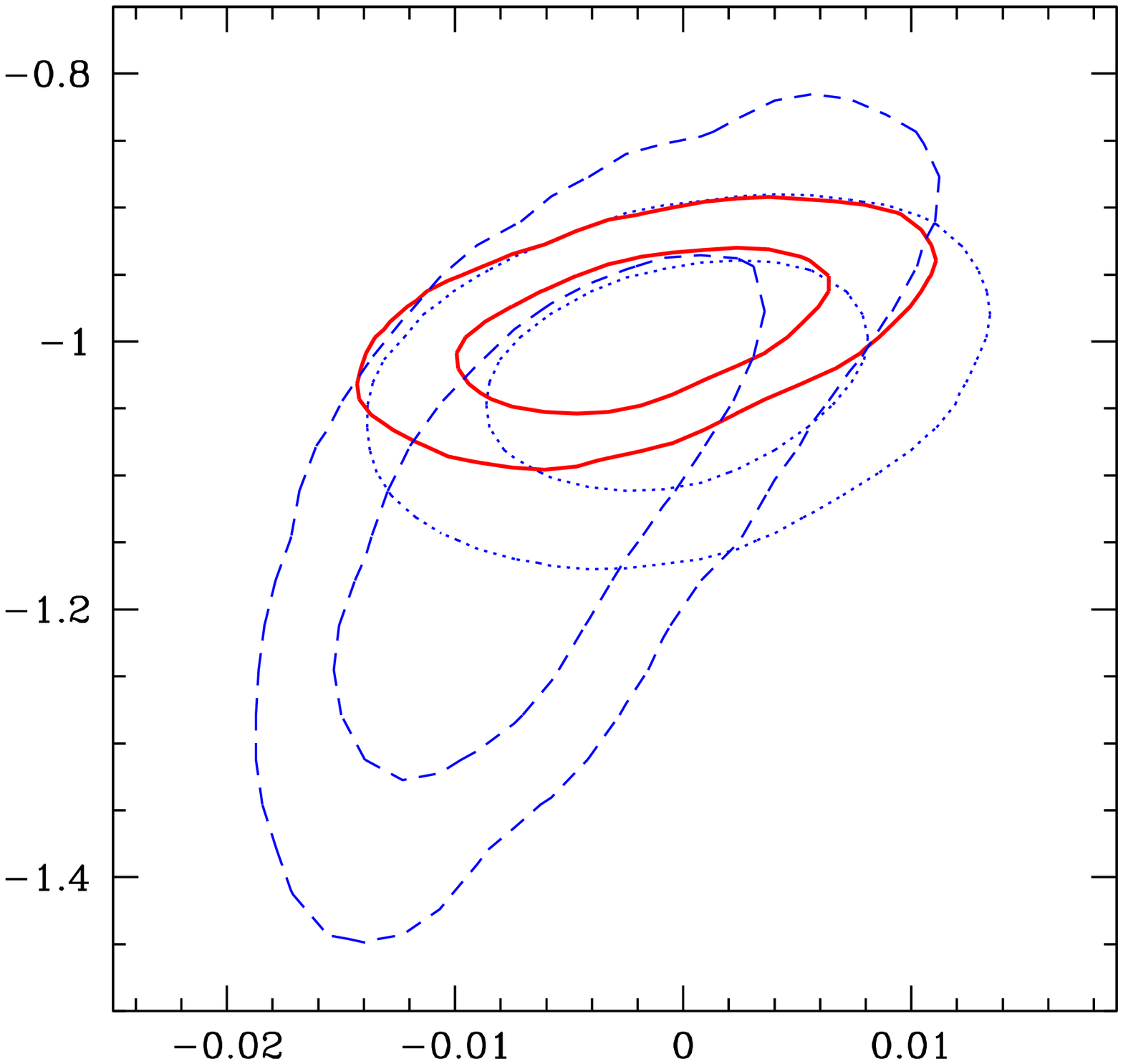}{$\Omega_k$}{$w$}
  \end{minipage}
  \caption{The constraint on the curvature of Universe and dark energy
    equation of state parameter in \emph{WCDM} model with $\Omega_k\ne
    0$.  Larger contours --- for \emph{WMAP7+BAO+$H_0$+SN} (dotted
    lines) and for \emph{WMAP7+BAO+$H_0$+CL} dataset (dashed lines),
    smaller contours --- for all the data
    (\emph{WMAP7+BAO+$H_0$+SN+CL}) combined. }
  \label{fig:okwsncl2}
\end{figure}

\subsection{Curved Universe}
\label{sec:curvedu}

The constraint on the curvature of Universe and dark energy equation
of state parameter in \emph{WCDM} model with $\Omega_k\ne 0$ are shown
in Fig.~\ref{fig:okwsncl} and \ref{fig:okwsncl2}. In model with
non-zero space curvature the cluster data allow to obtain only less
powerful constraint on $w$, as compared to supernovae data (cf.\ left
and right panels in Fig.~\ref{fig:okwsncl}). However, taking in
account systematic errors of supernovae data, the accuracy of $w$
measurement appears to be comparable. From Fig.~\ref{fig:okwsncl2} one
can see that the data on galaxy cluster mass function combined with
all other data, including the data on SN Ia, allow to significantly
improve the measurement of the parameter of dark energy equation of
state.

\subsection{Time-dependent equation of state}
\label{sec:w0wa}

\begin{figure}
  \centering
  \begin{minipage}{0.9\linewidth}
    \smfigure{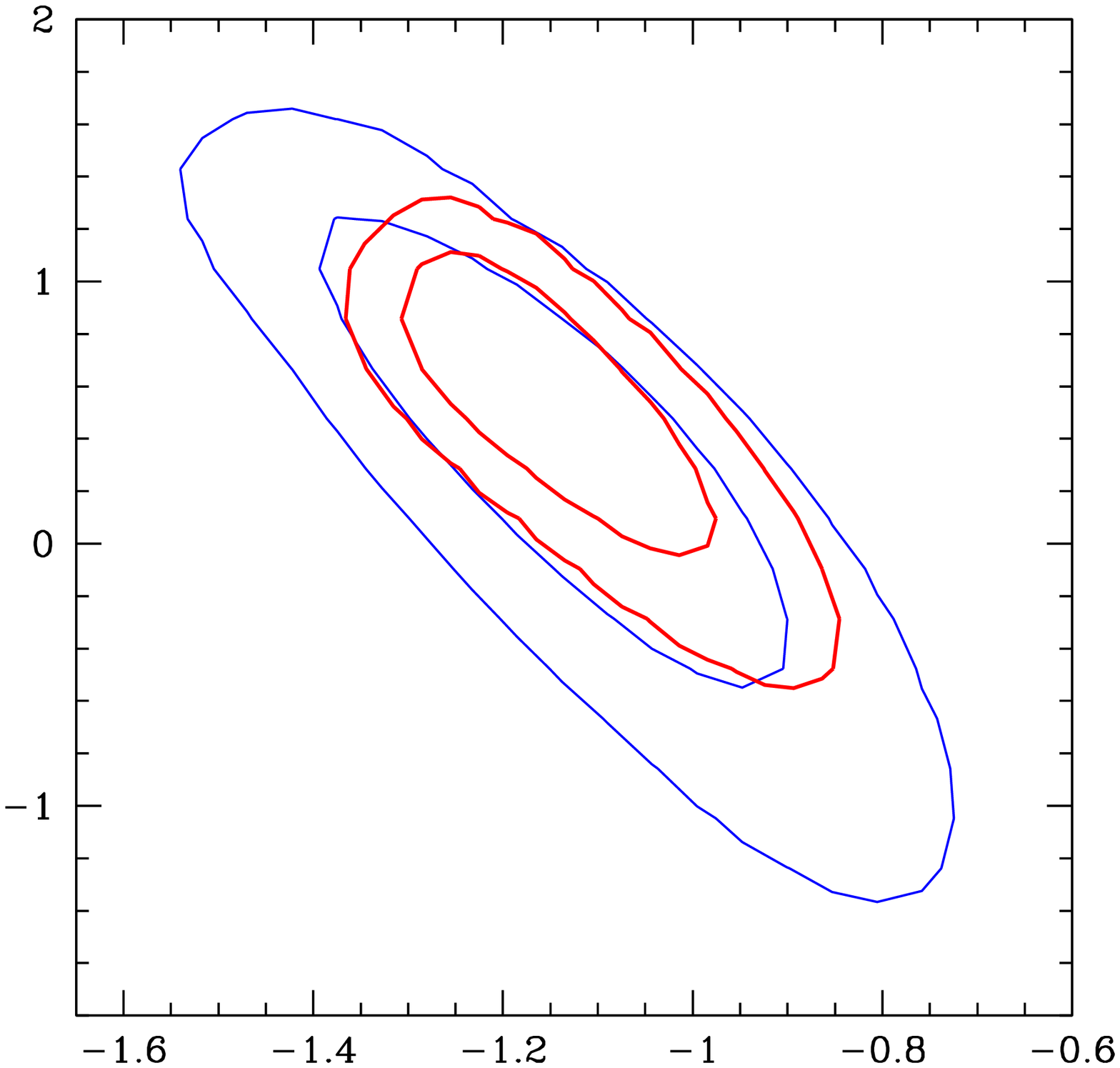}{$w_0$}{$w_a$}
  \end{minipage}
  \caption{The constraints on dark energy equation of state parameters
    in \emph{WACDM} model. Larger contours --- for
    \emph{WMAP7+BAO+$H_0$+SN} dataset, smaller contours --- for
    \emph{WMAP7+BAO+$H_0$+SN+CL} dataset.}
  \label{fig:w0wa}
\end{figure}

The constraints on the parameters of time-dependent dark energy
equation of state, defined as: $$w(a) = w_0 + w_a(1-a)~,$$ are shown
in Fig.~\ref{fig:w0wa} (see also Table~\ref{tab:pars}, \emph{WACDM}
model). In order to calculate CMB anisotropy power spectra we used
Parametrized Post-Friedmann approach \citep{fang08}, which was done
using the correspondent module for \texttt{CosmoMC} software.  One can
see, that galaxy cluster mass function data significantly improve the
overall constraints for $w_0$ and $w_a$ parameters. Therefore,
existent cluster mass function data give comparably powerful
constraints for these parameters, as compared to supernovae data.

\section{Summary}

In our work we present the results of detailed analysis of
cosmological parameter constraints which were obtained from
combination of galaxy cluster mass function measurements (Vikhlinin et
al., 2009a,b) with the other cosmological data, obtained recently. We
show, that this allow to significantly improve the constraints for
many cosmological parameters.

All considered cosmological data combined together are consistent with
the model of flat Universe with cosmological constant
(\emph{$\Lambda$CDM}). In frames of this model all considered
cosmological data, with galaxy cluster mass function data taken in
account, give the most powerful constraints on $\sigma_8$ and
$\Omega_m h^2$ parameters. The constraints on other parameters, such
as $\Omega_m$, $H_0$, $\Omega_b$, are also improved. The data on
galaxy cluster mass function allow to constrain these parameters with
as high as $\approx 1\%$ accuracy (see Table~\ref{tab:pars}). At the
same time, the systematical errors from the uncertainties of cluster
mass measurements appear to be comparable to statistical ones.

The data on galaxy cluster mass function allow to obtain new important
constraints on total neutrino mass $\Sigma m_\nu$ and effective number
of neutrino species $\neff$. When massive neutrinos or additional
number of neutrino species are added to \emph{$\Lambda$CDM} model, the
strongest constraints are obtained: $\Sigma m_\nu<0.32$\,eV and
$\neff<3.74$.  However, if both total neutrino mass and effective
number of neutrino species are considered as free parameters in
\emph{$\Lambda$CDM} model, i.e., the model with arbitrary number of
massive neutrinos is considered, the constraints turn out to be
significantly less strong: $\Sigma m_\nu<0.72$\,eV and $\neff<4.62$.
Moreover, all considered cosmological data are somewhat better fitted
by the model with non-zero neutrino mass $\Sigma m_\nu\approx 0.4$\,eV
and larger than standard value of the number of neutrino species,
$\neff\approx 4$.

These constraints can have an important application to study of the
possibility of the existence of light sterile neutrinos with masses
near $1$\,eV, which were suggested to explain the results of short
baseline neutrino oscillations searches experiments \emph{LSDN} and
\emph{MiniBooNE}, and also recently discovered, so-called reactor
neutrino anomaly.  However, as compared to cosmological constraints
obtained in our work, in order to explain the results of these
experiments, too large number of additional neutrino species (more
than one) or too large neutrino masses $\Sigma m_\nu\approx 1$\,eV are
required.

In our work the updated constraints on dark energy equation of state
parameters are also presented. The constraints obtained using galaxy
cluster mass function data, with no use of SN Ia measurements, are
considered as well. We show, that taking in account systematical
uncertainties, dark energy parameters constraints from cluster mass
function data are comparable in their power with those from SN Ia
observations. Moreover, in contrast to SN Ia data, the errors of dark
energy parameters measurements from current cluster mass function data
are mainly statistical, and are not dominated by systematic
uncertainties. Therefore, expansion of a sample of galaxy clusters,
suitable for accurate measurements of galaxy cluster mass function,
will allow to further improve dark energy constraints in future.

\acknowledgements

We are grateful to D.~S.~Gorbunov for useful discussion of the results
of our work and for a number of important remarks and suggestions. In
this work the results of calculations on MVS-100K supercomputer of
Joint Supercomputer Center of the Russian Academy of Sciences (JSCC
RAS) were used. The work is supported by Russian Foundation for Basic
Research, grants 08-02-00974, 09-02-12384-ofi-m, 10-02-01442,
11-02-12271-ofi-m, the Program for Support of Leading Scientific
Schools of the Russian Federation (Nsh-5603.2012.2), and the Programs
of the Russian Academy of Sciences P-19 and OPhN-16.

\end{document}